%% file: main.tex
\documentclass[sigconf]{acmart}

\usepackage{algorithm}
\usepackage{algorithmic}
\usepackage{amsmath}
\usepackage{booktabs}

\usepackage{adjustbox}
\usepackage{subcaption}
\usepackage{caption}
\usepackage{tabularx}
\usepackage{makecell}
\usepackage{xcolor}
\usepackage{pifont}
\usepackage{multirow} 
\usepackage{array}

\usepackage{cmap}
\usepackage[T1]{fontenc}
\AtBeginDocument{%
  }

\copyrightyear{2026}
\acmYear{2026}
\setcopyright{cc}
\setcctype{by-nc-nd}
\acmConference[ICCAD '26]{IEEE/ACM International Conference on Computer-Aided Design}{November 08--12, 2026}{San Jose, CA, USA}
\acmBooktitle{IEEE/ACM International Conference on Computer-Aided Design (ICCAD '26), November 08--12, 2026, San Jose, CA, USA}
\acmDOI{10.1145/3831252.3834223}
\acmISBN{979-8-4007-2873-0/2026/11}

\begin{document}

\title{SPIMOE: Exploiting Hybrid Sparsity for Reasoning MoE Inference on Heterogeneous PIM Architectures}


\titlenote{
This work was supported by the National Natural Science Foundation of China (Grant No.62572036).
}







\author{Rubing Yang$^1$, Cenlin Duan$^{2\dagger}$, Yingjie Qi$^1$, Xiaolin He$^1$, Xiao Ma$^1$, Jianlei Yang$^{1}$}
\authornote{Corresponding authors are \textit{Cenlin Duan and Jianlei Yang}, Email: \url{duancenlin@buaa.edu.cn, jianlei@buaa.edu.cn}.}
\affiliation{%
  \institution{$^1$School of Computer Science, Beihang University, Beijing, China}
  \country{}
}
\affiliation{%
  \institution{$^2$School of Integrated Circuits and Systems, Beihang University, Beijing, China}
  \country{}
}

\renewcommand{\shortauthors}{Rubing Yang, Cenlin Duan, Yingjie Qi, Xiaolin He, Xiao Ma, Jianlei Yang}

\renewcommand{\authors}{%
Rubing Yang, Cenlin Duan, Yingjie Qi, Xiaolin He, Xiao Ma, and Jianlei Yang%
}

\begin{abstract}

Long-reasoning Mixture-of-Experts (MoE) models expose two coupled inference bottlenecks: growing KV caches shift the critical path toward attention, while sparse expert activation causes load imbalance and low hardware utilization.
Although Processing-in-Memory (PIM) offers a promising way to mitigate data movement overhead, existing PIM-based accelerators typically optimize attention or FFNs in isolation.
We propose SPIMOE, the first co-design framework that exploits hybrid sparsity for efficient MoE inference on heterogeneous PIM architectures.
SPIMOE combines adaptive expert routing with block-sparse attention and physical KV-cache eviction, and disaggregates attention and FFNs across SRAM-PIM and HBM-PIM.
Static expert mapping and dynamic sub-batch scheduling further balance channel loads and overlap the two paths.
Evaluations show that SPIMOE achieves up to $8.35\times$ end-to-end speedup over an NVIDIA A100 GPU and $3.33\times$ speedup in MoE FFN execution over PIMoE, while preserving reasoning accuracy comparable to full-attention baselines.
\end{abstract}

\ccsdesc[500]{Computer systems organization~Heterogeneous (hybrid) systems}
\ccsdesc[300]{Hardware~Emerging architectures}
\ccsdesc[500]{Computing methodologies~Machine learning}

\keywords{Processing-in-Memory, Mixture-of-Experts, Sparse Attention, Adaptive Routing}


\maketitle

\input{tex/section1-introduction}

\input{tex/section2-background_motivation}

\input{tex/section3-method}

\input{tex/section4-experiment}
\input{tex/section5-conclusion}

\bibliographystyle{ACM-Reference-Format}
\bibliography{ref}

\newpage
\appendix









\end{document}

%% file: tex/section1-introduction.tex
\section{Introduction}\label{intro}

Large Language Models (LLMs) have achieved remarkable success across diverse domains, ranging from natural language understanding to autonomous code generation and scientific discovery~\cite{vaswani2017attention,brown2020language,achiam2023gpt,touvron2023llama}.
To further scale model capacity without proportionally increasing computation, Mixture-of-Experts (MoE) has emerged as a promising architecture by activating only a subset of experts for each token~\cite{fedus2022switch,lepikhin2020gshard,yang2025qwen3,li2025slimmoe,dai2024deepseekmoe}.
However, the growing adoption of Chain-of-Thought (CoT) reasoning shifts LLM inference from short-response generation to long-chain reasoning~\cite{guo2025deepseek,chen2025towards,wei2022chain,openai2024reason}, fundamentally changing the inference workload and exposing the limitations of conventional von Neumann-based accelerators.
First, prolonged reasoning sequences incur a linear growth of the KV cache footprint, pivoting the system's execution profile from being feed-forward network (FFN)-dominated to attention-dominated.
Meanwhile, the dynamic expert routing mechanism in MoE introduces irregular, unpredictable accesses to expert weights, further aggravating data movement overhead.
As a result, conventional accelerators struggle to efficiently support sparse long-reasoning MoE inference.

While Processing-in-Memory (PIM)~\cite{duan2024towards,duan2025efficient, qi2025ciminus} is promising for mitigating data movement overhead, directly mapping long-reasoning MoE workloads onto PIM architectures remains non-trivial.
As summarized in Tab.~\ref{tab:comparison_with_prior_works}, existing PIM-based accelerators~\cite{pan2025stratum,yun2024duplex,wu2025pimoe,tu2022trancim,fu2025h} mainly optimize either MoE or attention in isolation, and therefore fall short of holistically supporting long-reasoning workloads.
First, MoE-centric designs fail to exploit the temporal activation sparsity of experts for dynamic pruning, leading to redundant data movement. 
Second, current attention-centric optimizations often incur significant accuracy degradation in long-chain reasoning tasks and fail to alleviate the fundamental KV cache capacity bottleneck. 
Finally, the lack of Attention-FFN decoupling and holistic co-optimization results in severe resource underutilization and an inability to adapt to the dynamic critical path shifts inherent in multi-step reasoning.

\begin{table}[t]
\centering
\caption{Comparison with prior PIM-based LLM acceleration works.}
\label{tab:comparison_with_prior_works}

\renewcommand{\arraystretch}{1.0}
\setlength{\tabcolsep}{2pt}

\scriptsize
\begin{tabular}{l p{1.6cm} p{1.65cm} c c c c}
\toprule
\textbf{Work} & \textbf{Architecture} & \textbf{Algorithm} & \textbf{\makecell{Focus \\ Layer}} & \textbf{\makecell{Reason- \\ ing}} & \textbf{\makecell{Co- \\ Optim.}} & \textbf{\makecell{KV \\ Drop}} \\
\midrule
Stratum~\cite{pan2025stratum} & NPU+DRAM-PIM & Topic Classify & MoE & $\times$ & $\times$ & $\times$ \\
Duplex~\cite{yun2024duplex} & xPU+DRAM-PIM & None & MoE & $\times$ & $\times$ & $\times$ \\
PIMoE~\cite{wu2025pimoe} & NPU+DRAM-PIM & N:M Pruning & MoE & $\times$ & $\checkmark$ & $\times$ \\
TranCIM~\cite{tu2022trancim} & SRAM-PIM & Sparse Attn. & Attn. & $\times$ & $\checkmark$ & $\checkmark$ \\
H2EAL~\cite{fu2025h} & Hybrid-Bonding & Sparse Attn. & Attn. & $\times$ & $\checkmark$ & $\times$ \\
\textbf{Ours} & \textbf{Hetero. PIM} & \textbf{\makecell[l]{Adaptive Routing \\ + Sparse Attn.}} & \textbf{MoE+Attn.} & $\checkmark$ & $\checkmark$ & $\checkmark$ \\
\bottomrule
\end{tabular}
\vspace{-8pt}
\end{table}

To bridge this gap, we propose SPIMOE, the first hybrid sparse heterogeneous PIM framework tailored for long-reasoning MoE inference. 
By orchestrating algorithm-hardware co-design, SPIMOE improves the efficiency of both Attention and expert-FFN execution throughout the entire reasoning process.
Our key contributions are as follows:

\begin{figure}[!t]
  \centering
  \includegraphics[width=0.8\columnwidth]{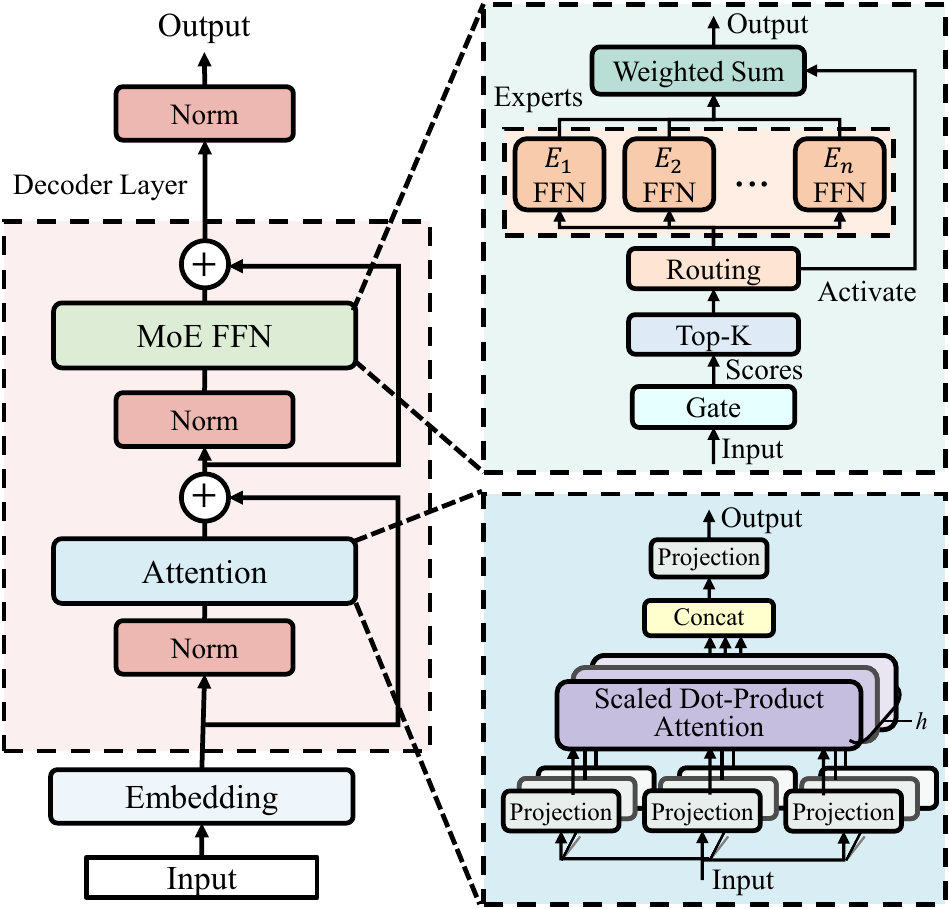}
  \caption{Model structure of MoE transformer.}
  \label{fig:moe_structure}
  \vspace{-20pt}
\end{figure}

\begin{enumerate}
    \item \textbf{Algorithm Level: }We propose \textbf{reasoning-aware and PIM-friendly} techniques, specifically a block-sparse attention mechanism and an adaptive expert routing strategy. 
    These techniques integrate sparse block selection with physical KV cache eviction to prune redundant blocks at semantic CoT boundaries,
    and utilizes adaptive expert routing with dynamic thresholds and thinking-critical expert boosting to balance HBM-PIM channel workloads.
    \item \textbf{Hardware Level:} We design a heterogeneous SRAM-PIM and HBM-PIM system with decoupled attention/FFN data paths.
    We further introduce static expert mapping and dynamic sub-batch scheduling to alleviate expert load imbalance and interconnect bottleneck.
    
    \item We build an end-to-end system-level evaluation framework for SPIMOE. Compared with an NVIDIA A100 GPU baseline, SPIMOE achieves up to \textbf{8.35$\times$} speedup at batch size 8 and \textbf{6.27$\times$} speedup at sequence length 2K. Compared with the prior PIMoE design, SPIMOE delivers \textbf{3.33$\times$} speedup for expert FFN execution.
\end{enumerate}

%% file: tex/section2-background_motivation.tex
\section{Background and Motivations}

\subsection{MoE Architecture}\label{2.1}

MoE has emerged as a representative sparse scaling architecture in modern LLMs~\cite{fedus2022switch,lepikhin2020gshard,cai2025survey}.
As illustrated in Fig.~\ref{fig:moe_structure}, each MoE layer comprises a router and multiple expert FFNs.
A lightweight router selects the top-k experts for each token, and only the corresponding expert FFNs are executed. 
This token-wise sparse activation mechanism improves model scalability, but it also makes expert computation highly dependent on routing decisions.
As LLM applications increasingly shift from short-response generation to long-chain reasoning, such routing-dependent execution exhibits substantially different runtime characteristics.

\begin{figure}[t]

  
  \begin{minipage}[t]{0.47\columnwidth}
    \centering
    \includegraphics[width=\linewidth]{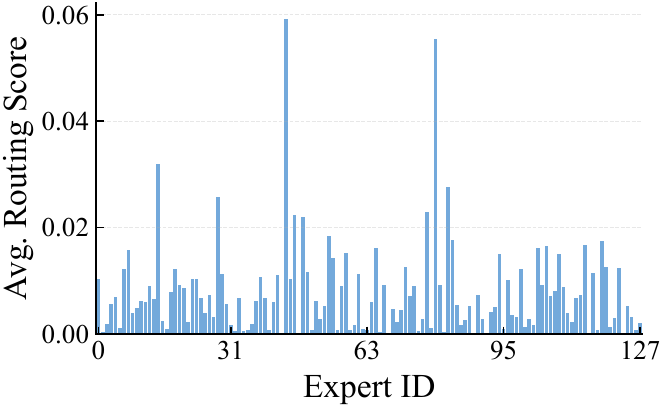}
    \caption{Average routing score distribution across experts for thinking-critical tokens.}
    \label{fig:motivation_expert_weight_think}
  \end{minipage}
  \hfill
  \begin{minipage}[t]{0.47\columnwidth}
    \centering
    \includegraphics[width=\linewidth]{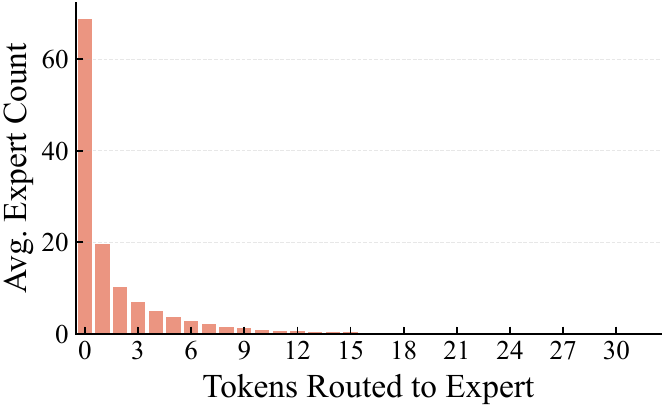}
    \caption{Distribution of tokens routed to each expert per decode step with batch size~32.}
    \label{fig:motivation_expert_avg_act_token}
  \end{minipage}

    \begin{minipage}[t]{0.95\columnwidth}
      \centering
      \includegraphics[width=0.85\linewidth]{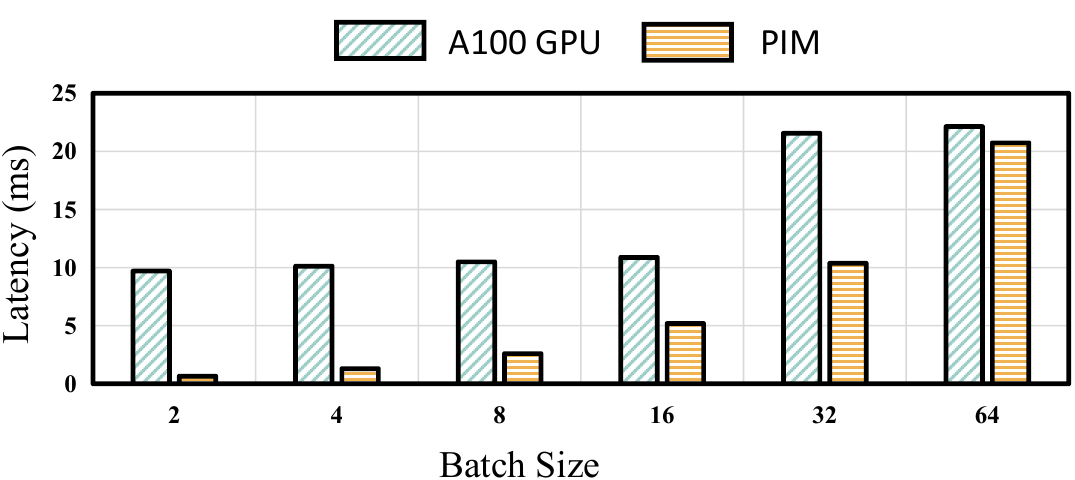}
      \caption{MoE FFN execution time comparison under enforced identical routing decisions across tokens.}
      \label{fig:motivation_ffn_gpu_pim}
    \end{minipage}

  \vspace{-16pt}
\end{figure}

During prolonged decoding, the continuous growth of the KV cache makes attention increasingly dominant in runtime, as it becomes more constrained by storage capacity and memory accesses.
Meanwhile, expert activation varies across tokens, layers, and reasoning steps, leading to stronger sparsity, irregular expert-weight accesses, and dynamic load imbalance. 
As shown in Fig.~\ref{fig:motivation_expert_weight_think}, the average routing scores of reasoning-critical tokens are highly skewed across experts, indicating that only a small subset of experts consistently receives high importance. 
Moreover, Fig.~\ref{fig:motivation_expert_avg_act_token} shows a clear long-tail distribution in the number of tokens routed to each expert at every decode step.
Most experts receive only a few tokens or are not activated at all. 
These observations reveal strong temporal sparsity and severe expert load imbalance in long-reasoning MoE inference.
Taken together, they suggest that long-reasoning MoE inference is increasingly constrained by KV cache-intensive attention and irregular expert execution, rather than raw compute alone. 
As a result, such workloads are difficult to accelerate efficiently on conventional architectures.


\subsection{PIM Architecture}
PIM has emerged as a promising approach to alleviating the memory wall by integrating computation into memory, making it particularly suitable for memory-bound workloads~\cite{park2024attacc,seo2024ianus,heo2024neupims}.
This property benefits both major components of long-reasoning MoE inference.
For attention, as discussed in Sec.~\ref{2.1}, long-chain reasoning continuously enlarges the KV cache, which naturally favors near-data execution.
For expert FFNs, long-reasoning MoE exhibits stronger sparsity and a more long-tailed expert activation pattern. 
As a result, expert-FFN execution typically appears as a small-batch, narrow GEMM or GEMV workload.
Such workloads are difficult to execute efficiently on GPUs, as GPU architectures are primarily optimized for larger and denser matrix operations, as illustrated in Fig.~\ref{fig:motivation_ffn_gpu_pim}.
Although existing PIM-based studies have demonstrated the potential of near-data execution for MoE~\cite{pan2025stratum,yun2024duplex,wu2025pimoe} and sparse workloads~\cite{tu2022trancim,fan2025sparse,fu2025h}, most of them do not explicitly target long-reasoning MoE inference.
Moreover, the reliance on homogeneous execution substrates prevents prior designs from efficiently accommodating the divergent bottlenecks of attention and expert FFN under long-reasoning MoE inference.
%

\subsection{Motivations of SPIMOE}
The above analysis reveals three key design requirements.
First, attention and expert FFNs exhibit different access patterns and bottlenecks, favoring a heterogeneous PIM architecture with decoupled data paths.
Second, highly skewed and long-tailed expert activation patterns result in severe load imbalance and redundant expert execution. 
Such characteristics are difficult to address through hardware scheduling alone, thereby motivating joint support for architecture-level load balancing and algorithm-level adaptive expert routing.
Third, KV cache growth makes attention increasingly dominant, motivating reasoning-aware sparse attention that reduces both computation and physical cache capacity without degrading reasoning quality.
\textbf{These observations collectively motivate a hybrid heterogeneous PIM framework with joint algorithm-hardware optimization for long-reasoning MoE inference.}

%% file: tex/section3-method.tex
\section{SPIMOE Framework}
\label{sec:framework}

\subsection{Framework Overview}
\label{subsec:overview}


\begin{figure}[t]
  \centering
  \includegraphics[width=0.9\columnwidth]{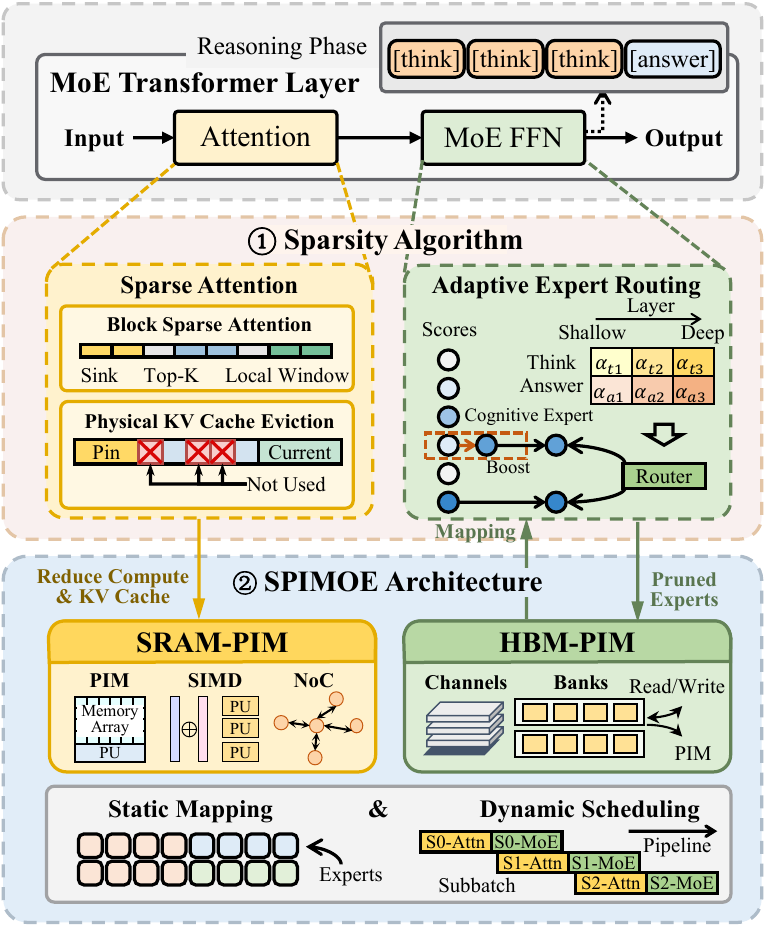}
  \caption{Overview of SPIMOE framework.}
  \label{fig:method_overview}
  \vspace{-12pt}
\end{figure}

SPIMOE is an algorithm-architecture co-design framework that maps long-reasoning MoE inference onto a heterogeneous PIM, as shown in Fig.~\ref{fig:method_overview}. 
It adopts Attention-FFN Disaggregation (AFD): SRAM-PIM executes Attention, while HBM-PIM handles QKV projections and expert FFNs.
Adaptive expert routing dynamically prunes low-contribution experts based on the reasoning phase and layer depth, reducing FFN computation and mitigating channel congestion in HBM-PIM. Block sparse attention combined with reasoning-aware KV cache eviction reduces per-step attention cost and reclaims capacity for longer reasoning sequences.
To further improve utilization, SPIMOE adopts a dynamic sub-batch scheduling strategy that partitions the batch into smaller sub-batches, allowing attention and MoE FFN to execute in an overlapped pipeline across consecutive sub-batches.

\subsection{Sparsity Algorithms}
\label{subsec:sparsity}

\subsubsection{Adaptive Expert Routing Algorithm}
\label{subsubsec:adaptive_expert_routing}

\mbox{}\par
\begin{figure}[t]
  \centering
  \begin{subfigure}[t]{0.49\columnwidth}
    \centering
    \includegraphics[width=\linewidth]{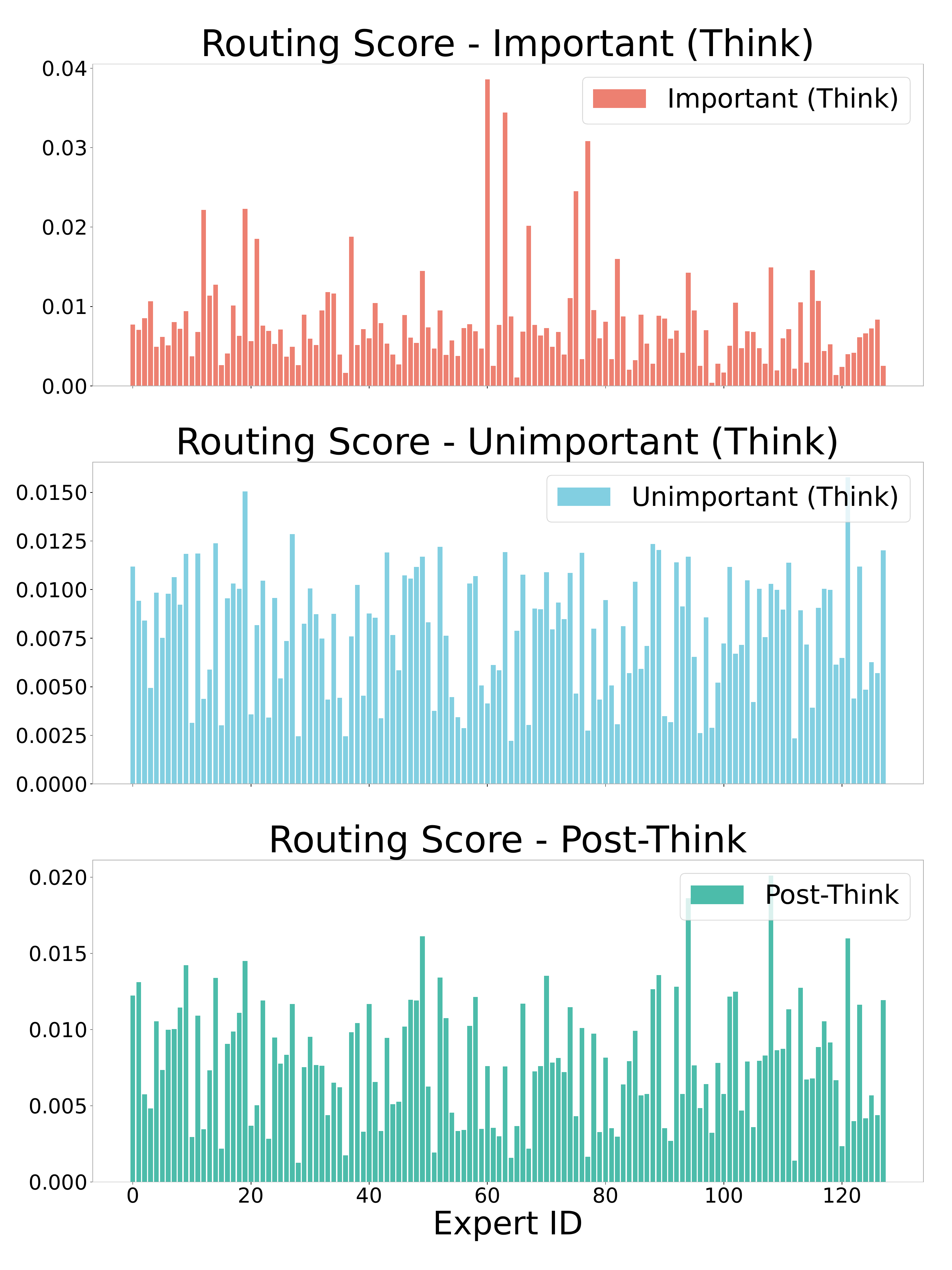}
    \caption{First layer.}
    \label{fig:section3_routing_score_layer0}
  \end{subfigure}
  \hfill
  \begin{subfigure}[t]{0.49\columnwidth}
    \centering
    \includegraphics[width=\linewidth]{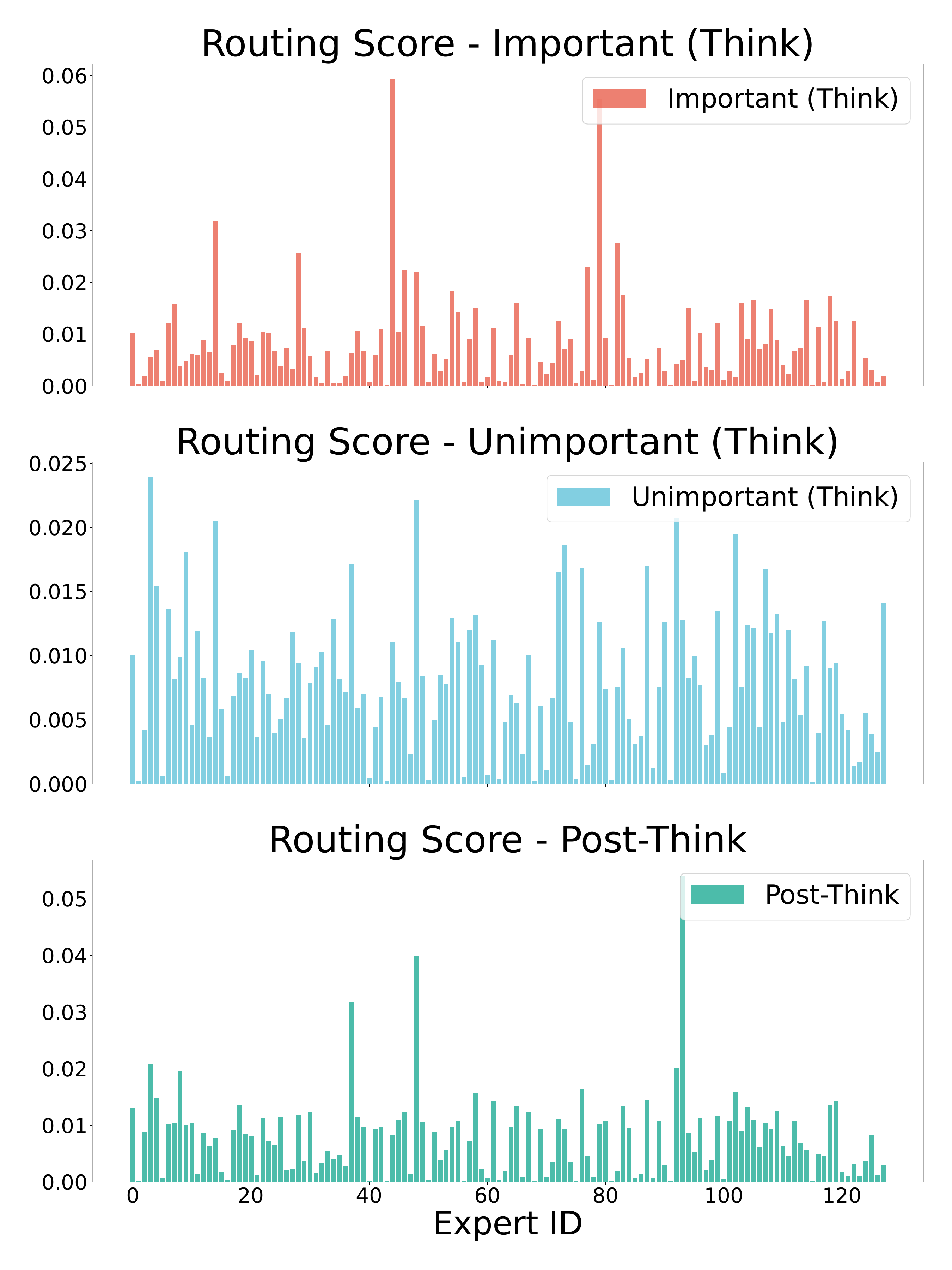}
    \caption{Last layer.}
    \label{fig:section3_routing_score_layer47}
  \end{subfigure}
  \caption{Routing score distribution across experts. Each subplot shows important (think), unimportant (think), and post-think routing scores.}
  \label{fig:section3_routing_score}
  \vspace{-16pt}
\end{figure}

\noindent\textbf{Routing Weight Analysis.}
Fig.~\ref{fig:section3_routing_score} shows the routing score distribution across experts at different layers. The distributions are highly skewed, with a few experts dominating, indicating substantial sparsity. This skewness increases with depth, as the distributions in deeper layers become more concentrated. The distributions also vary across reasoning phases, with routing concentrating on a few dominant experts during the answer phase. Notably, important reasoning tokens tend to focus on a small subset of critical experts, a phenomenon also observed in prior work~\cite{wang2025two}, where these experts are referred to as \textit{Cognitive Experts}. These observations motivate our adaptive routing strategy.

\noindent\textbf{Phase-Depth Alpha Partitioning.}
We define a pruning coefficient $\alpha_{p,r}$, where $p \in \{\textit{think}, \textit{answer}\}$ denotes the reasoning phase detected online via the \textit{</think>} token, and $r \in \{\textit{shallow}, \textit{middle}, \textit{deep}\}$ denotes the layer depth tier. At layer $l\in \{1,\ldots,L\}$ where $L$ denotes the total
number of layers, only experts whose routing weight satisfies
\begin{equation}
  w_i^{(l)} \geq \alpha_{p,r} \cdot \max\nolimits_{j \in \text{Top-}K} w_j^{(l)}
  \label{eq:alpha_threshold}
\end{equation}
are retained, with a minimum count $k_{\min}$ enforced. The answer phase permits higher $\alpha$ values due to its more concentrated weight distribution.

\noindent\textbf{Cognitive Expert Boost.}
For each layer $l$, we identify the top-$N_{\mathrm{cog}}$ cognitive experts $\mathcal{E}^{\mathrm{cog}}_l$ through offline profiling of normalized activation frequency on reasoning-critical tokens~\cite{wang2025two} from a calibration set disjoint from evaluation data. 
At inference time, their routing weights are amplified by a boost factor $\beta > 1$ prior to the threshold test:
\begin{equation}
  \tilde{w}_i^{(l)} =
  \begin{cases}
    \beta \cdot w_i^{(l)}, & i \in \mathcal{E}^{\mathrm{cog}}_l, \\
    w_i^{(l)}, & \text{otherwise}.
  \end{cases}
  \label{eq:cognitive_boost}
\end{equation}
The boosted weights $\tilde{w}_i^{(l)}$ then undergo Eq.~\eqref{eq:alpha_threshold}, ensuring cognitive experts are preferentially retained under aggressive pruning.

\begin{algorithm}[t]
\caption{Adaptive expert routing.}
\label{alg:adaptive_expert_routing}
\small
\begin{algorithmic}[1]
\REQUIRE Routing weights $\{w_i^{(l)}\}$, phase $p$, depth tier $r$, cognitive expert set $\mathcal{E}^{\mathrm{cog}}_l$, expert mapping $\mathcal{P}$, boost $\beta$, thresholds $\{\alpha_{p,r}\}$, minimum $k_{\min}$
\ENSURE Pruned expert set $\mathcal{S}_{prune}$ and renormalized weights
\FOR{each expert $i$ in Top-$K$}
  \STATE $\tilde{w}_i^{(l)} \leftarrow \beta \cdot w_i^{(l)}$ if $i \in \mathcal{E}^{\mathrm{cog}}_l$ and $p = \textit{think}$, else $w_i^{(l)}$
\ENDFOR
\STATE $\tau \leftarrow \alpha_{p,r} \cdot \max_j \tilde{w}_j^{(l)}$
\STATE $\mathcal{S}_{prune} \leftarrow \{i : \tilde{w}_i^{(l)} \geq \tau\}$; pad to $k_{\min}$ if $|\mathcal{S}_{prune}| < k_{\min}$
\STATE Compute per-channel load $\eta_c$ from $\mathcal{P}$
\FOR{each $i \in \mathcal{S}_{prune}$ on congested channel ($\eta_c > \mu_\eta + \sigma_\eta$)}
  \IF{$i \notin \mathcal{E}^{\mathrm{cog}}_l$ \textbf{and} $|\mathcal{S}_{prune}| > k_{\min}$}
    \STATE $\mathcal{S}_{prune} \leftarrow \mathcal{S}_{prune} - \{i\}$
  \ENDIF
\ENDFOR
\STATE Renormalize weights of remaining experts in $\mathcal{S}_{prune}$
\RETURN $\mathcal{S}_{prune}$
\end{algorithmic}
\end{algorithm}

\noindent\textbf{Parameter Calibration.} The phase-depth thresholds \(\{\alpha_{p,r}\}\), boost factor \(\beta\), and $k_{min}$ are calibrated offline on a small calibration set disjoint from the evaluation data. 
A lightweight grid search selects configurations on the accuracy-performance Pareto frontier.
The selected parameters are reused across datasets for the same model, while new model families require one-time recalibration due to different routing distributions.

\noindent\textbf{PIM-Aware Expert Pruning.}
On HBM-PIM, experts on the same channel create serialized DRAM computation and NoC communication bottlenecks. Let $a_{l,e} \in \{0,1\}$ denote the activation indicator for expert $(l,e)$. The instantaneous load per channel is $\eta_c = \sum_{(l,e) \in \mathcal{P}(c)} a_{l,e}$, counting the number of currently active experts mapped to channel $c$. We prioritize pruning experts $i \notin \mathcal{E}^{\mathrm{cog}}_l$ on congested channels ($\eta_c > \mu_\eta + \sigma_\eta$), as shown in Algorithm~\ref{alg:adaptive_expert_routing}.

\subsubsection{Sparse Attention Algorithm}
\label{subsubsec:sparse_attn}

\mbox{}\par
\begin{figure}[t]
  \centering
  \begin{subfigure}[t]{0.49\columnwidth}
    \centering
    \includegraphics[width=0.8\linewidth]{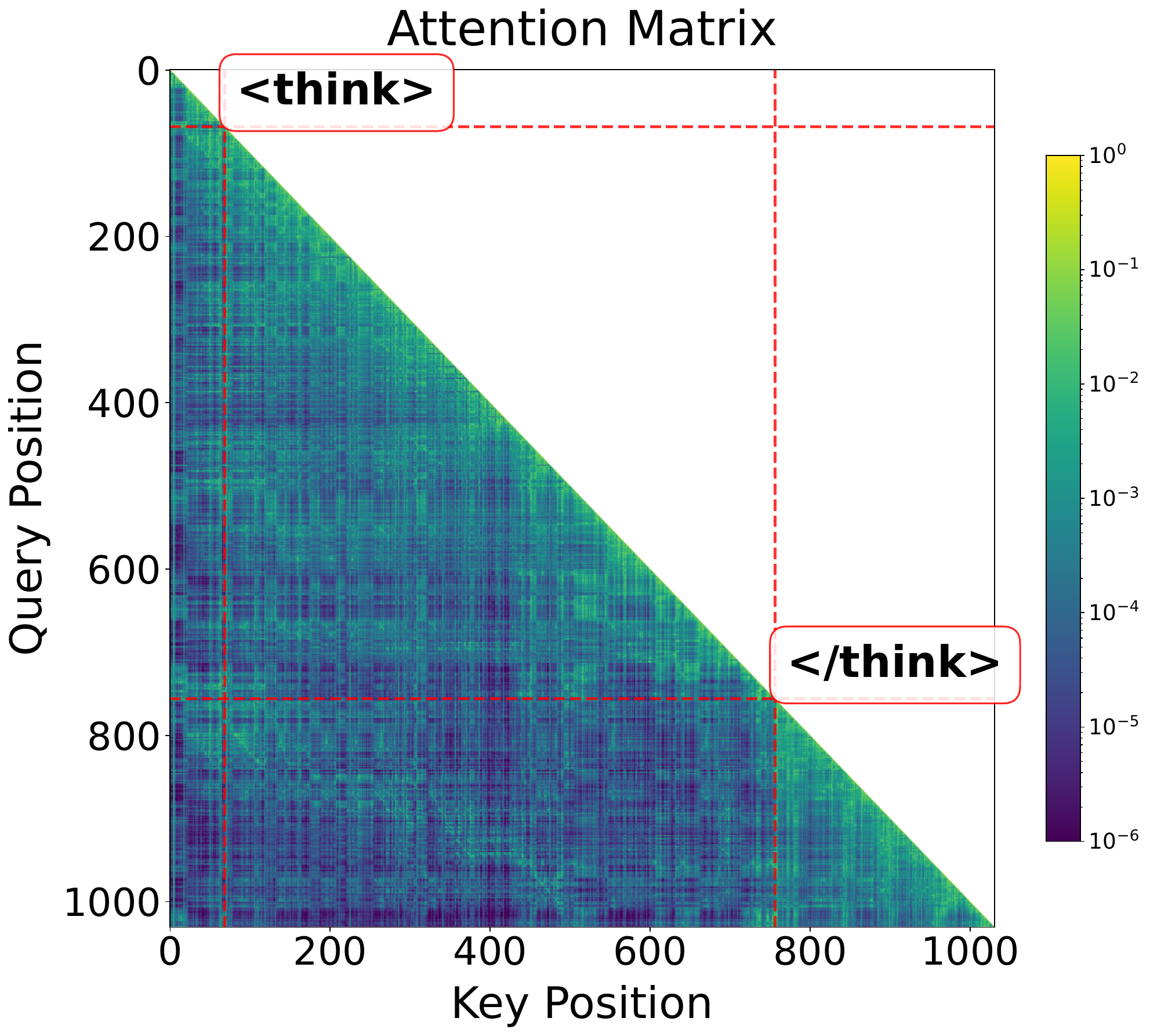}
    \caption{Middle layer.}
    \label{fig:sec3_attention_heatmap_layer0}
  \end{subfigure}
  \hfill
  \begin{subfigure}[t]{0.49\columnwidth}
    \centering
    \includegraphics[width=0.8\linewidth]{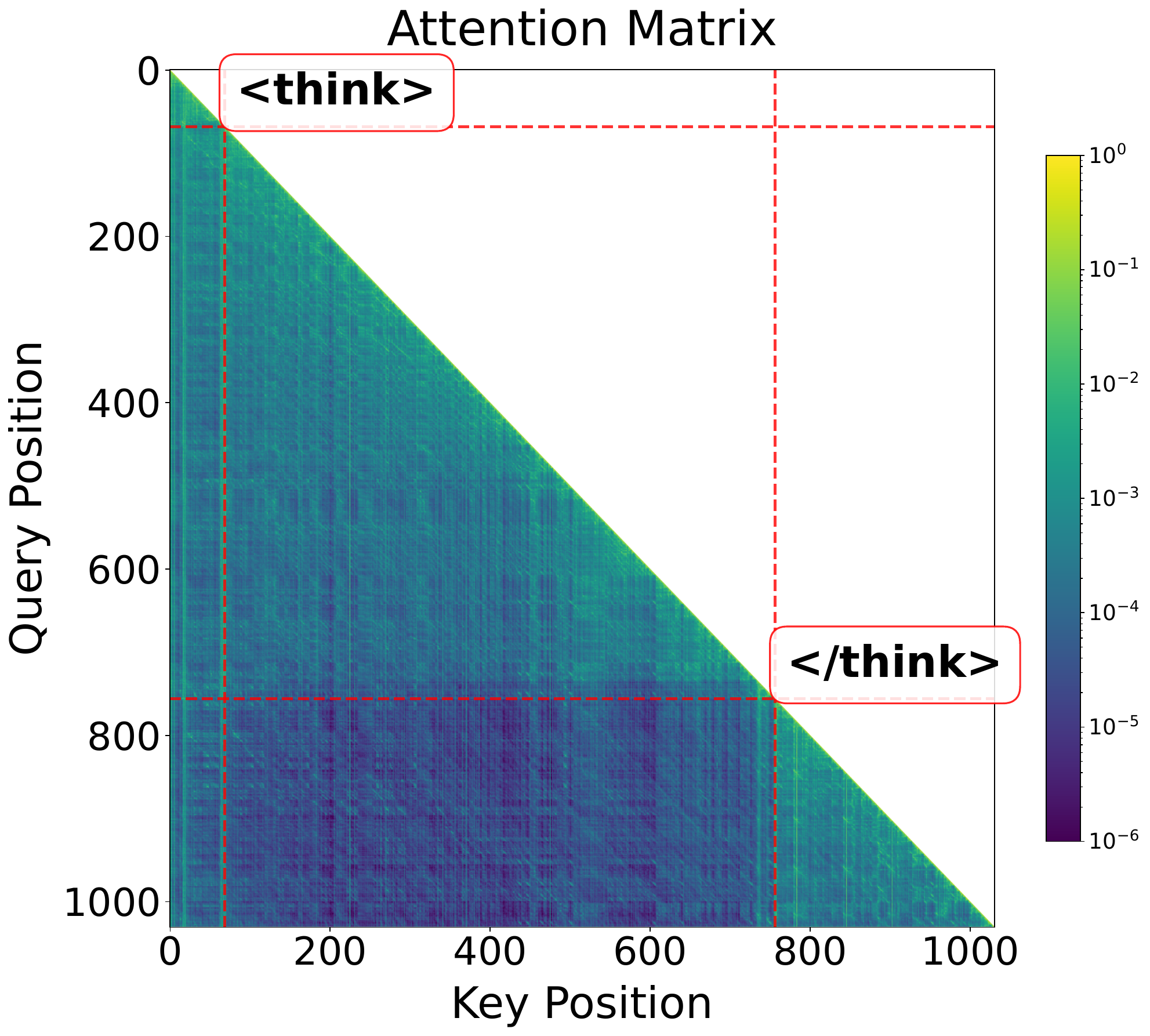}
    \caption{Last layer.}
    \label{fig:sec3_attention_heatmap_layer47}
  \end{subfigure}
  \caption{Attention heatmaps during reasoning at different layers.}
  \label{fig:sec3_attention_heatmap}
\end{figure}

\noindent\textbf{Attention Pattern Analysis.} Fig.~\ref{fig:sec3_attention_heatmap} reveals a three-stage evolution of attention patterns across network depth. In middle layers, the model selectively attends to a small subset of reasoning blocks visible as vertical stripes, while aggregation behavior emerges as summary blocks consolidate information from earlier segments. In deep layers, attention further collapses to a more local pattern with only a narrow local window and a few globally important anchor tokens. Once the thinking phase concludes, most reasoning tokens quickly become less important. This progressive sparsification indicates that the reasoning chain serves as an intermediate result rather than persistent memory, motivating both block-granularity sparse computation and physical eviction of obsolete KV entries.

\begin{figure}[t]
  \centering
  \includegraphics[width=0.9\columnwidth]{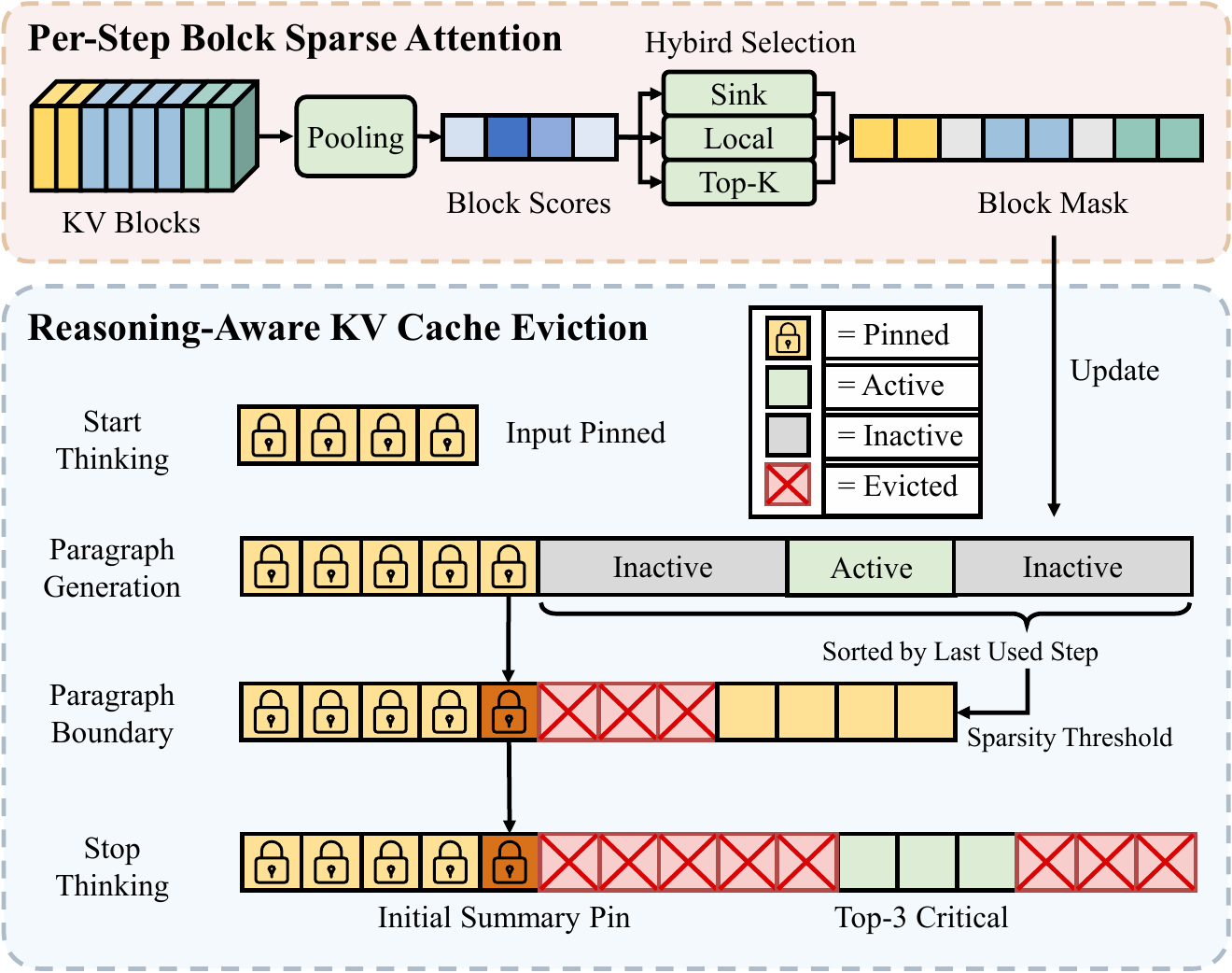}
  \caption{Design of sparse attention algorithm.}
  \label{fig:sparse_attn_algorithm}
  \vspace{-8pt}
\end{figure}

Existing methods, such as Quest~\cite{tang2024quest} and MInference~\cite{jiang2024minference} reduce computation via token skipping or mask-based sparsity, but the skipped KV entries remain physically resident. As illustrated in Fig.~\ref{fig:sparse_attn_algorithm}, we propose a two-level scheme: block sparse attention for per-step computation reduction, and reasoning-aware KV cache eviction for storage savings.

\noindent\textbf{Block Sparse Attention.} The KV sequence of length $s$ is partitioned into chunks of size $B_{chunk}$ aligned with the SRAM macro row width. A representative key per chunk is obtained via mean pooling: $\bar{k}_j = B_{chunk}^{-1} \sum_{t \in \text{chunk}_j} k_t$. At each decode step, the query $q$ computes coarse scores $s_j = q^\top \bar{k}_j$ against all $N = \lceil s/B_{chunk} \rceil$ chunk representatives, and the visible KV cache for attention is formed as:
\begin{equation}
  \mathcal{V} = \underbrace{\mathcal{V}_{\text{sink}}}_{n_s \text{ tokens}} \cup \underbrace{\mathcal{V}_{\text{local}}}_{n_l \text{ tokens}} \cup \underbrace{\mathcal{V}_{\text{Top-K}}}_{K_b \text{ chunks by } s_j},
  \label{eq:visible_kv}
\end{equation}
where $\mathcal{V}_{\text{sink}}$ preserves prompt context as persistent global anchors, $\mathcal{V}_{\text{local}}$ maintains generation coherence, and $\mathcal{V}_{\text{Top-K}}$ captures the salient reasoning blocks. The attention output is computed exclusively over $\mathcal{V}$:
\begin{equation}
  o = \text{Softmax}\!\left(q K_\mathcal{V}^\top / \sqrt{d_h}\right) V_\mathcal{V},
  \label{eq:sparse_attn}
\end{equation}
reducing per-step complexity from $O(s)$ to $O(\rho s)$ under sparsity ratio $\rho$.

\noindent\textbf{Physical KV Cache Eviction.}
Block sparse attention reduces computation but not physical cache size. To achieve genuine storage savings, we introduce physical eviction triggered at semantic boundaries.
A designated anchor layer records per-chunk last-used timestamps as a cross-layer importance proxy. 
Eviction is triggered at structural boundaries such as paragraph separators \texttt{\textbackslash n\textbackslash n} and the end-of-thinking marker \textit{</think>}.
If no boundary is detected for 2,048 consecutive tokens and KV cache occupancy exceeds 80\%, a conservative fallback trigger is used.

Upon triggering, the cache is partitioned into three regions. \textbf{Pinned} chunks include prompt tokens and important chunks of the initial paragraph, which are permanently retained. \textbf{Current} chunks belong to the ongoing segment and are excluded from eviction. \textbf{Previous} chunks are historical eviction candidates sorted by last-used timestamps; only the top $\rho_{\text{evict}}$ fraction of most recently attended chunks is retained. Evicted entries are physically removed across all layers.
Evicted entries are reclaimed at block granularity and returned to a global free list, avoiding sub-page fragmentation; incremental chunk updates with reduction every \(B_{chunk}\) tokens incur \(O(1/B_{chunk})\) amortized overhead.
Unlike mask-based approaches that merely reduce dynamic energy, physical eviction directly reclaims scarce capacity, enabling longer reasoning within the same hardware budget.


\subsection{SPIMOE Architecture}
\label{subsec:architecture}

\subsubsection{Overall Architecture}
\label{subsubsec:overall_arch}

\mbox{}\par
\begin{figure}[t]
  \centering
  \includegraphics[width=0.9\columnwidth]{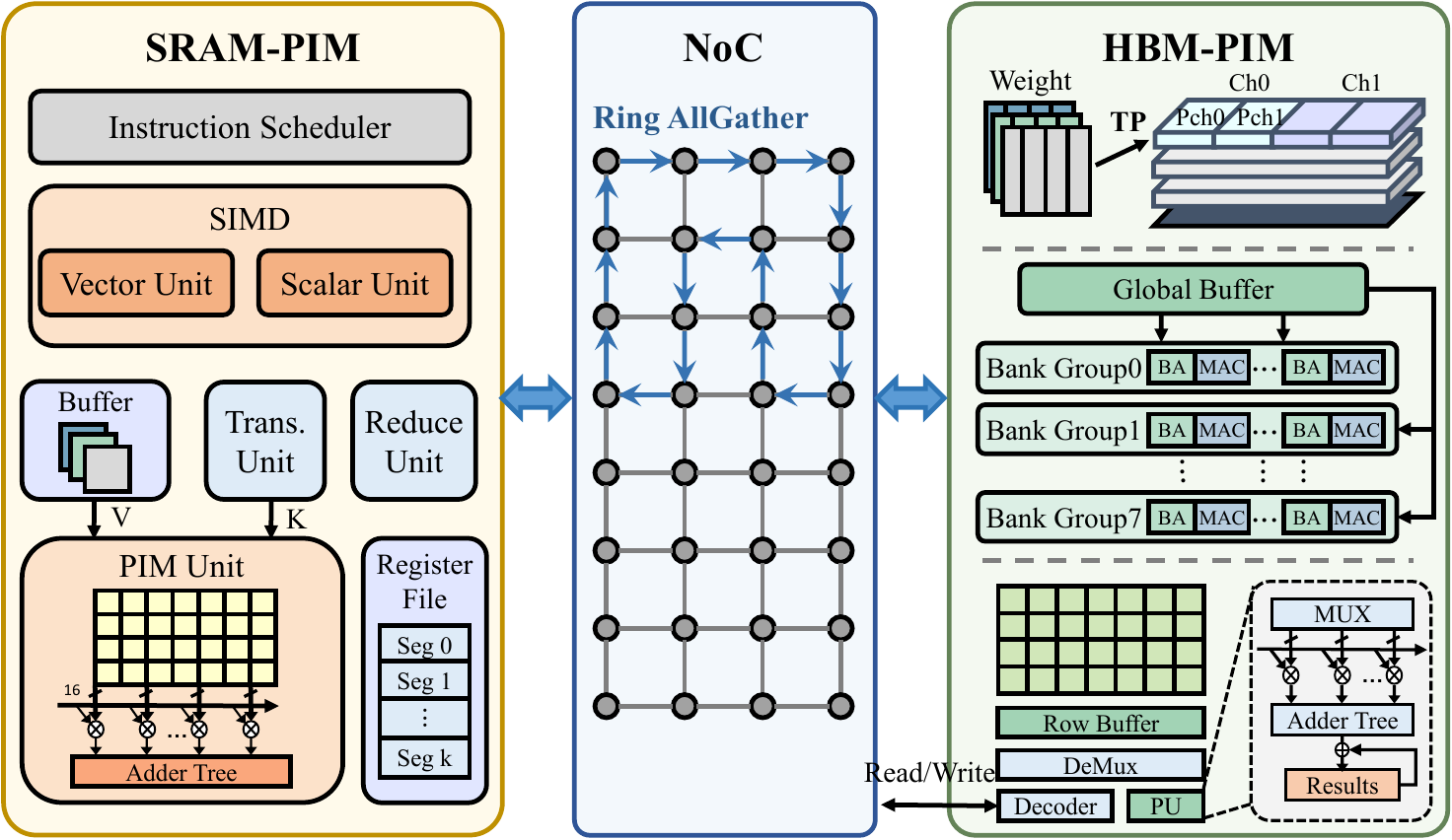}
  \caption{Overall architecture of SPIMOE.}
  \label{fig:architecture}
  \vspace{-15pt}
\end{figure}

\noindent
The fundamental design principle is Attention-FFN Disaggregation, as shown in Fig.~\ref{fig:architecture}. The two dominant operator classes in MoE transformer inference are mapped onto physically distinct PIM substrates whose characteristics match the respective computational profiles.

\noindent\textbf{SRAM-PIM.}
The SRAM-PIM subsystem contains 32 independent cores optimized for decode-phase attention. Each core integrates a PIM unit for matrix-vector multiplication, a vector compute unit for nonlinear functions such as softmax, a scalar compute unit for scalar operations, a transpose unit for K cache transposition, and a reduce unit for Top-K computation supporting expert routing and block-sparse attention selection. The low-latency, configurable SRAM substrate is well-suited for the irregular, access-intensive attention computation pattern.

\noindent\textbf{HBM-PIM.}
The HBM-PIM subsystem adopts the memory architecture design from Newton~\cite{he2020newton}, while the in-situ MAC units integrated in each bank perform GEMV operations on resident weight matrices without external data movement. In compute mode, the input activation vector is broadcast via a global buffer to all banks in parallel, exploiting the massive internal bandwidth of the TSV-interconnected HBM3 stack. This approach follows the characteristics of MoE inference, where each activated expert processes a few tokens on average, resulting in an operation that maps to the bank-parallel PIM execution model.

\noindent\textbf{System Topology.}
HBM channels and SRAM cores are connected through a $4 \times 8$ NoC mesh running at 256\,bits/flit and 800\,MHz, with a channel-to-core mapping. Each NoC link supports single-directional communication. The per-layer dataflow alternates between HBM-PIM and SRAM-PIM, enabling pipelined overlap across consecutive sub-batches as detailed in Section~\ref{subsubsec:scheduling}.
\begin{figure}[t]
  \centering
  \includegraphics[width=0.85\columnwidth]{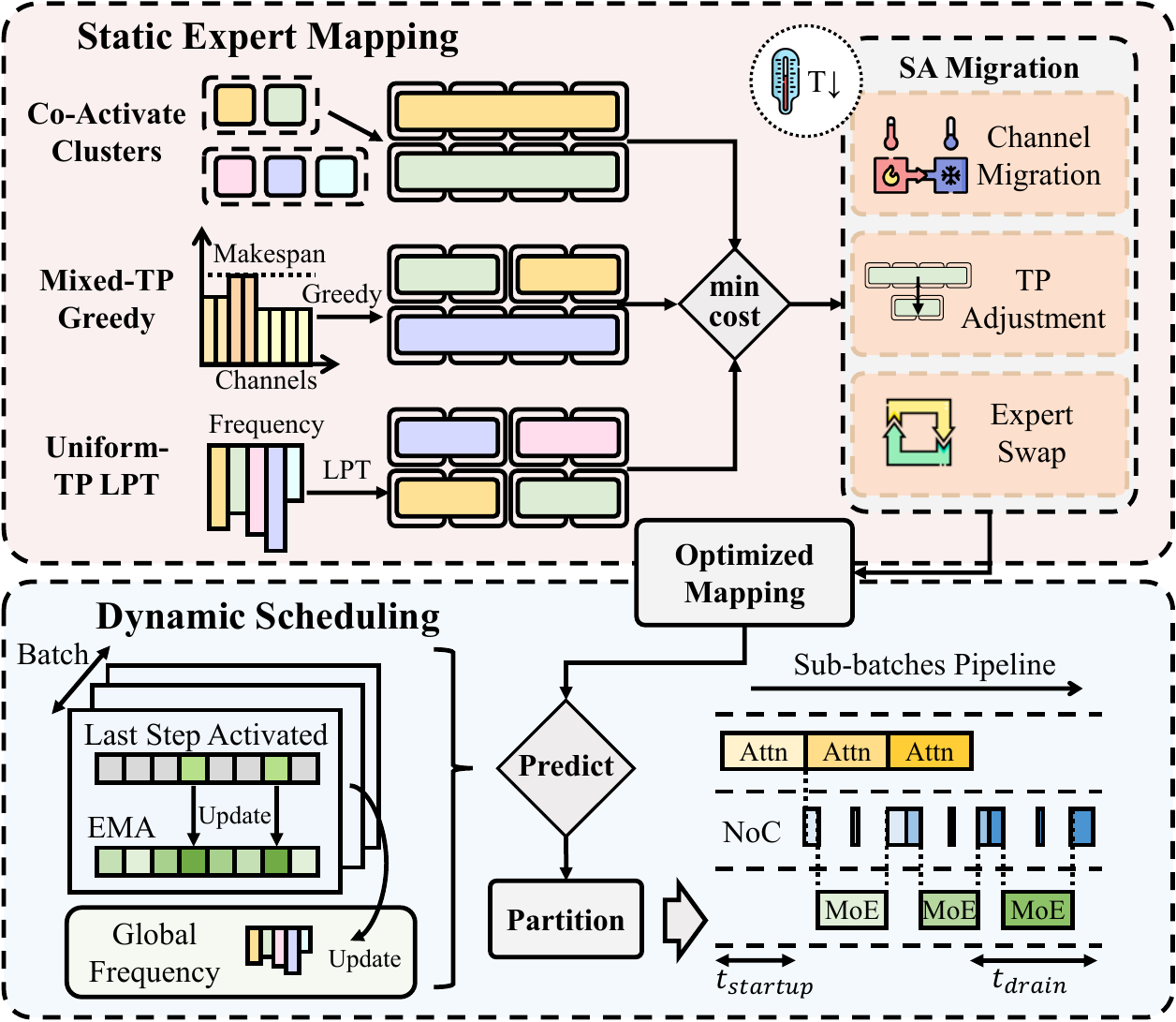}
  \caption{Design of mapping and scheduling strategy.}
  \label{fig:static_mapping_dynamic_schedue}
  \vspace{-15pt}
\end{figure}

\subsubsection{Dataflow and Communication Design}
\label{subsubsec:dataflow}

\mbox{}\par\noindent\textbf{Per-Layer Operator Pipeline.}
Each transformer layer is decomposed into 17 operator stages executed alternately on SRAM-PIM and HBM-PIM, with five NoC communication phases interleaved: KV all-gather before attention, activation all-gather before projection and before FFN1, TP communication around FFN2, and MoE output gather after FFN2.

Two modes are supported depending on how the FFN2 weight matrix is partitioned. In \emph{row mode}, each TP shard produces a partial output that is summed via all-reduce after FFN2 ($2(t{-}1)$ ring rounds, data size $d_{\text{size}}$). In \emph{col mode}, each shard requires the full intermediate activation, necessitating an all-gather before FFN2 ($t{-}1$ rounds, data size $d_{\text{inter}}/n_{\text{act}}$). The mode is selected based on the model's intermediate-to-model dimension ratio.

After FFN2, expert results are aggregated in three sub-phases: (1)~TP all-reduce within each expert's shard group (row mode only); (2)~inter-expert all-gather across the first cores of all activated experts; (3)~result scatter to distribute the final output.

\noindent\textbf{Ring Topology Optimization.}
Ring all-gather and all-reduce latency is dominated by the maximum bidirectional hop count in the ring. Given $n$ participating cores on the $4 \times 8$ mesh, we find a Hamiltonian cycle with ring order $\pi = (\pi_1,...,\pi_n)$ that minimizes $\max_i [\text{hop}(\pi_i, \pi_{i+1}) + \text{hop}(\pi_{i+1}, \pi_i)]$, where $\pi_{n+1}=\pi_1$. For rectangular sub-meshes, a snake-order traversal yields optimal rings in $O(n)$ time; for general core sets, a greedy nearest-neighbor construction followed by 2-opt local search is applied. Optimized ring orders are cached per core set to avoid recomputation.

\subsection{Mapping and Scheduling Mechanism}

\subsubsection{Static Expert Mapping}
\label{subsubsec:mapping}

\mbox{}\par\noindent
The static assignment of experts to HBM channels determines load balance and communication cost. An Integer Linear Programming (ILP) formulation over all layers, experts, channels, and TP factors yields an intractable search space; we instead adopt a multi-candidate heuristic strategy with simulated annealing refinement, as shown in Fig.~\ref{fig:static_mapping_dynamic_schedue}.

\noindent\textbf{Co-activation Clustering.}
From profiled activation data, we compute the co-activation ratio $\gamma_{ij} = \text{co\_act}(i,j) / \min(\text{cnt}_i, \text{cnt}_j)$ for each expert pair, where $\text{co\_act}(i,j)$ is the co-activation count of experts $i$ and $j$, and $\text{cnt}_i$ is the activation count of expert $i$
. Experts frequently activated by the same tokens are grouped into cliques via greedy maximal-clique extraction. 
Placing clique members on non-overlapping channel groups improves compute load balancing and enables parallel execution. 
The clique size determines the TP factor and channel group assignment for each member.

\noindent\textbf{Multi-Candidate Competition.}
For each layer, three candidate mapping strategies are evaluated: (1)~co-activation clustering with topology-aware channel assignment; (2)~mixed-TP greedy, where each expert independently selects the (TP, channel group) pair minimizing its cost; and (3)~uniform-TP with Longest Processing Time First (LPT) load balancing across channel groups. The candidate with the lowest per-step cost $\mathcal{J} = \mathcal{J}_{\text{DRAM}} + \mathcal{J}_{\text{NoC}}$ is selected, where $\mathcal{J}_{\text{DRAM}}$ is the channel computation makespan and $\mathcal{J}_{\text{NoC}}$ is the total NoC communication cost.

\noindent\textbf{SA Migration.}
The selected mapping is further refined via Simulated Annealing (SA) with three neighborhood operators: expert migration to another channel group, TP factor adjustment, and expert swap. The cost function is evaluated per decode step using profiled activation patterns, with incremental NoC updates for affected tokens only.

\subsubsection{Dynamic Scheduling}
\label{subsubsec:scheduling}

\mbox{}\par\noindent
Static expert mapping optimizes the average case but cannot adapt to runtime token-to-expert assignments. Dynamic sub-batch scheduling partitions the batch so that each sub-batch's MoE execution time approximates its attention time, improving pipeline efficiency, as shown in Fig.~\ref{fig:static_mapping_dynamic_schedue}.

\noindent\textbf{Cost Prediction.}
Since gate routing results are unavailable until after QKV projection, MoE cost must be predicted before attention begins.
At step~0, a lookup table provides cost estimates from offline profiling.
At subsequent steps, the predictor constructs a gating score matrix $\mathbf{G} \in \mathbb{R}^{L \times N_E}$:
\begin{equation}
  \mathbf{G} = \mathbf{\lambda}_{\text{prev}} \odot \mathbf{M}_{\text{prev}} + \mathbf{\lambda}_{\text{ema}} \odot \hat{\mathbf{A}} + \mathbf{\lambda}_{\text{freq}} \odot \mathbf{F},
  \label{eq:cost_fusion}
\end{equation}
where $\mathbf{M}_{\text{prev}}$ is the previous step's activation mask, $\hat{\mathbf{A}}$ is an Exponential Moving Average (EMA) of per-token expert affinity, and $\mathbf{F}$ is a global frequency prior. 
The blending coefficients \(\lambda_{\mathrm{prev}}\), \(\lambda_{\mathrm{ema}}\), and \(\lambda_{\mathrm{freq}}\) weight $\mathbf{M}_{\text{prev}}$, $\hat{\mathbf{A}}$, and $\mathbf{F}$, respectively, and are calibrated per layer from measured step-to-step autocorrelation.
The top-$K$ experts are selected from $\mathbf{G}$, and the predicted MoE cost is obtained via the expert-to-core mapping table.

\noindent\textbf{Sub-batch Partition.}
Sub-batches are formed via greedy token packing and closed when the accumulated predicted MoE cost of tokens in the sub-batch exceeds the attention time threshold determined by the sub-batch size.

\noindent\textbf{Pipeline Execution.}
An overlap execution mode interleaves attention and MoE across sub-batches: SRAM-PIM computes attention for sub-batch $bi+1$ while HBM-PIM executes MoE for sub-batch $bi$. The overall step latency is:
\begin{equation}
  T_{\text{step}} = T_{\text{startup}} + \sum_{bi} \max\!\left(T_{\text{attn}}^{(bi)},\; T_{\text{moe}}^{(bi)}\right) + T_{\text{drain}},
  \label{eq:pipeline}
\end{equation}
where $T_{\text{startup}}$ and $T_{\text{drain}}$ account for the first and last sub-batch that cannot be overlapped.

%% file: tex/section4-experiment.tex
\section{Experiment}
\label{sec:experiment}

\subsection{Experimental Setup}
\label{subsec:setup}

\noindent\textbf{Models.}
Qwen3-30B-A3B~\cite{yang2025qwen3} and Phi-mini-MoE-instruct~\cite{li2025slimmoe} are used for end-to-end evaluation. For component-level comparison with PIMoE, we use Switch-Large-128~\cite{fedus2022switch} and Switch-Base-16.
We additionally use the dense Qwen3-1.7B model for sparse-attention evaluation.
In Table~\ref{tab:experiment_model_config}, $L$ denotes the number of layers, $d$ denotes the model hidden size, and $d_{\text{inter}}$ denotes the total intermediate size of the FFN module. $H_Q$ and $H_{KV}$ denote the number of query heads and key-value heads, respectively, and $d_h$ is the hidden dimension per head. $N_E$ denotes the total number of experts, and $K$ denotes the number of activated experts per token.

\begin{table}[t]
\centering
\caption{Model configurations.}
\label{tab:experiment_model_config}
\renewcommand{\arraystretch}{1.0}
\setlength{\tabcolsep}{3pt}
\small 
\begin{tabular}{l c c c c c c c c c}
\toprule
\textbf{Model} & \textbf{\makecell{Params}} & \textbf{\makecell{$L$}} & \textbf{\makecell{$d$}} & \textbf{\makecell{$d_{\text{inter}}$}} & \textbf{\makecell{$N_E$}} & \textbf{\makecell{$K$}} & \textbf{\makecell{$H_Q$}} & \textbf{\makecell{$H_{KV}$}} & \textbf{\makecell{$d_h$}} \\
\midrule
Qwen3-30B-A3B          & 30B & 48 & 2048 & 6144 & 128 & 8 & 32 & 4  & 128 \\
Phi-mini-MoE           & 15B & 32 & 4096 &  960 &  16 & 2 & 32 & 8  & 128 \\
Switch-Large-128       & 24B & 12 & 1024 & 4096 & 128 & 1 & 16 & 16 &  64 \\
Switch-Base-16         & 0.9B &  6 &  768 & 3072 &  16 & 1 & 12 & 12 &  64 \\
\bottomrule
\end{tabular}
\end{table}

\noindent\textbf{Benchmarks.}
We evaluate on five reasoning benchmarks with varying difficulty levels.
For mathematical reasoning, GSM8K~\cite{cobbe2021gsm8k} covers grade-school arithmetic, MATH-500~\cite{math500hendrycks2021measuringmathematicalproblemsolving} covers high-school competition problems, and AIME 2024~\cite{aime} covers competition-level challenges. For science reasoning, GPQA~\cite{rein2023gpqagraduatelevelgoogleproofqa} covers graduate-level questions, while ARC-Challenge~\cite{clark2018arc} covers grade-school science logic. We enable reasoning for all accuracy evaluations.

\noindent\textbf{Hardware Specification.}
The hardware configuration of the proposed SPIMOE architecture is detailed in Tab.~\ref{tab:hardware_config}.
The system integrates $32$ SRAM-PIM cores with $4$ HBM-PIM modules, interconnected via a 2.5D silicon interposer. 
Each SRAM-PIM core operates at $800$~MHz, while each HBM-PIM bank integrates a $400$~MHz PU delivering $6.4$~GFLOPS for FP16 GEMV. The four HBM3 modules provide a total capacity of $96$~GB, sufficient to accommodate MoE models such as Qwen3-30B-A3B.

\begin{table}[t]
\centering
\caption{Hardware configuration.}
\label{tab:hardware_config}
\renewcommand{\arraystretch}{1.1}
\small
\begin{tabular}{|m{0.15\columnwidth}|m{0.2\columnwidth}|m{0.5\columnwidth}|}
\hline
SPIMOE   & Composition & 32 SRAM-PIM core, 4 HBM-PIM \\
\hline
\multirow[l]{3}{*}{SRAM-PIM} & \makecell[l]{Memory \\ Configuration} & 384KB Activation Memory, 32KB Temp Memory \\
\cline{2-3}
 & PIM Unit & 800MHz, 16 Macro Groups, 16 PUs per MG, 12.8 GFLOPS per PU \\
\hline
\multirow[l]{3}{*}{HBM-PIM}  &  \makecell[l]{Memory \\ Configuration} & HBM3, 24GB/HBM, 8 dies, 8 DRAM per die, 2 Channel per DRAM, 2 Pseudo Channels per Channel, 4 Bank Groups per pCH, 8 Banks per BG \\
\cline{2-3}
 & Processing Unit (PU) & 400MHz, 1 PU per Bank, 6.4 GFLOPS per PU \\
\hline
\end{tabular}
\end{table}

\noindent\textbf{Baselines.}
We use an NVIDIA A100-80GB GPU running standard inference as the baseline. For sparse attention, we compare against two representative methods, MInference~\cite{jiang2024minference} and Quest~\cite{tang2024quest}. 
For MoE acceleration on PIM architectures, we compare with PIMoE~\cite{wu2025pimoe} at the MoE FFN component level, since PIMoE does not support attention or KV cache management.

\noindent\textbf{Simulation Infrastructure.}
Our simulation infrastructure builds upon the framework developed for our prior heterogeneous PIM work, HPIM~\cite{duan2026hpim}, and is extended to model SPIMOE-specific mechanisms. 
The SRAM-PIM subsystem is modeled using our in-house CIMFlow framework~\cite{qi2025cimflow}, with digital modules implemented in Verilog HDL and synthesized using Synopsys Design Compiler at 12~nm for area and power estimation. 
The HBM-PIM subsystem follows the bank-level PIM organization of Newton~\cite{he2020newton} and is modeled using an extended DRAMsim3~\cite{li2020dramsim3} under the HBM3 specification~\cite{HBM3}; its PUs and global buffers are also synthesized at 12~nm. The $4\times8$ mesh NoC is modeled using an extended Noxim~\cite{catania2015noxim}.


\subsection{Performance Evaluation}
\label{subsec:perf}

\begin{figure}[t]
  \centering
  \begin{subfigure}[t]{0.49\columnwidth}
    \centering
    \includegraphics[width=\linewidth]{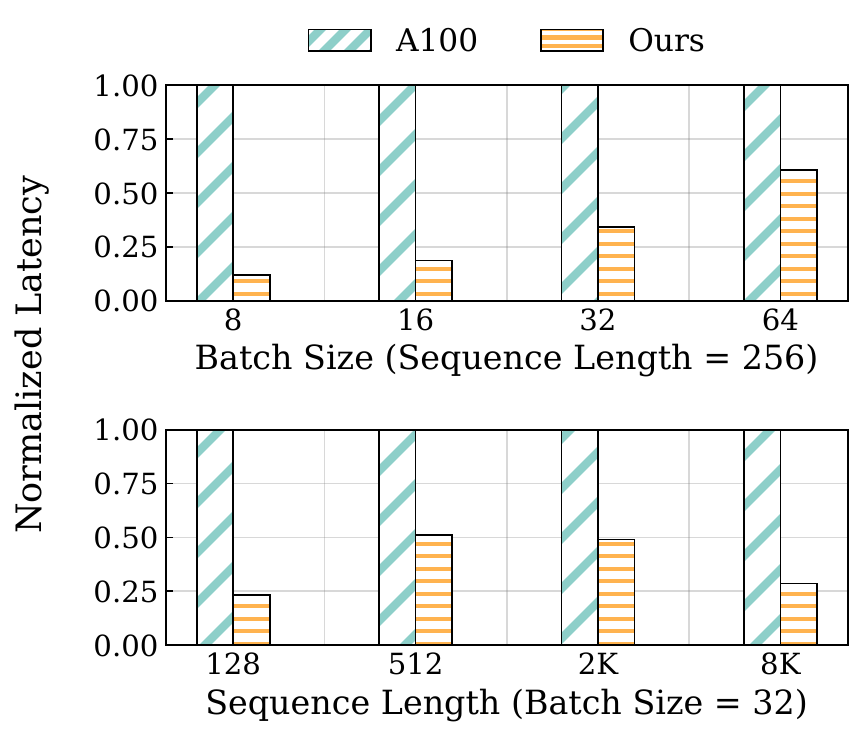}
    \caption{Qwen3-30B-A3B.}
    \label{fig:sec5_speedup_qwen}
  \end{subfigure}\hfill
  \begin{subfigure}[t]{0.49\columnwidth}
    \centering
    \includegraphics[width=\linewidth]{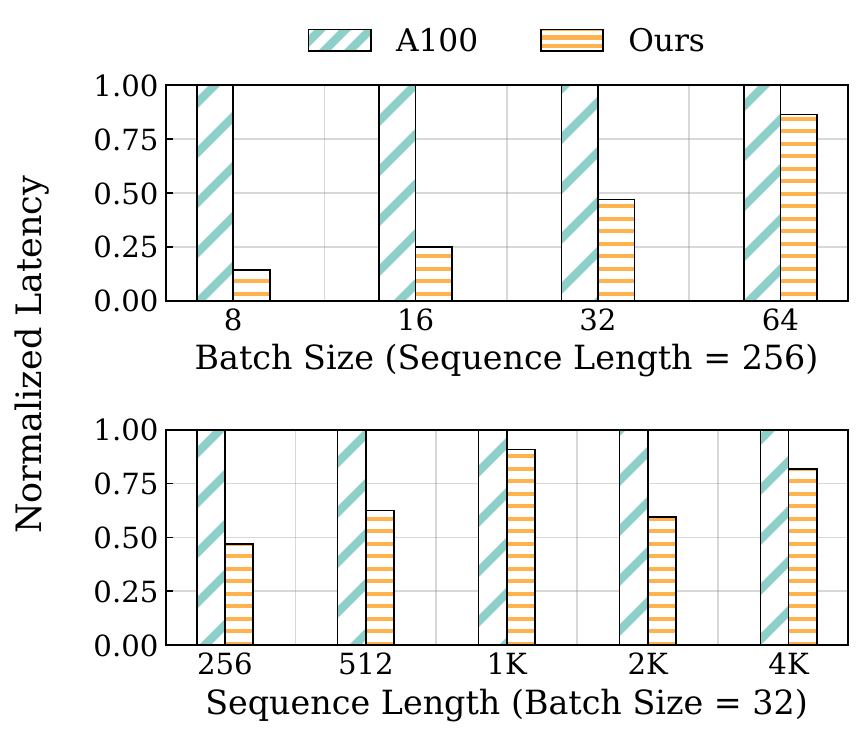}
    \caption{Phi-mini-MoE-instruct.}
    \label{fig:sec5_speedup_phi}
  \end{subfigure}
  \vspace{-6pt}
  \caption{Normalized latency vs.\ A100 GPU baseline.}
  \label{fig:sec5_speedup}
  \vspace{-8pt}
\end{figure}

\noindent\textbf{End-to-End Speedup.} Fig.~\ref{fig:sec5_speedup} presents the end-to-end speedup under two sweep dimensions. At small batch sizes where MoE FFN dominates the execution time, the heterogeneous PIM architecture achieves substantial acceleration: at batch size 8 with sequence length 256, Qwen3-30B-A3B achieves $8.35\times$ speedup, and Phi-mini-MoE-instruct achieves $7.08\times$ speedup over the A100 baseline. As batch size increases, the per-expert token count grows, and GPU GEMM utilization improves, reducing the performance gap with the GPU baseline; at batch size 64, the speedup diminishes as the workload transitions from bandwidth-bound to compute-bound.
Meanwhile, the sequence-length sweeps in Fig.~\ref{fig:sec5_speedup} show that, at a fixed batch size of 32, speedup initially decreases as the sequence length increases due to the growing attention cost.
It recovers at longer sequences as block-sparse attention reduces the SRAM-PIM workload. 
For Qwen3-30B-A3B at 8K tokens, SPIMOE achieves a $3.51\times$ speedup, demonstrating the effectiveness of the proposed co-design for long-reasoning workloads. 

\label{subsec:pimoe_cmp}


\begin{figure}[t]
  \centering
  \begin{subfigure}[t]{0.49\columnwidth}
    \centering
    \includegraphics[width=\linewidth]{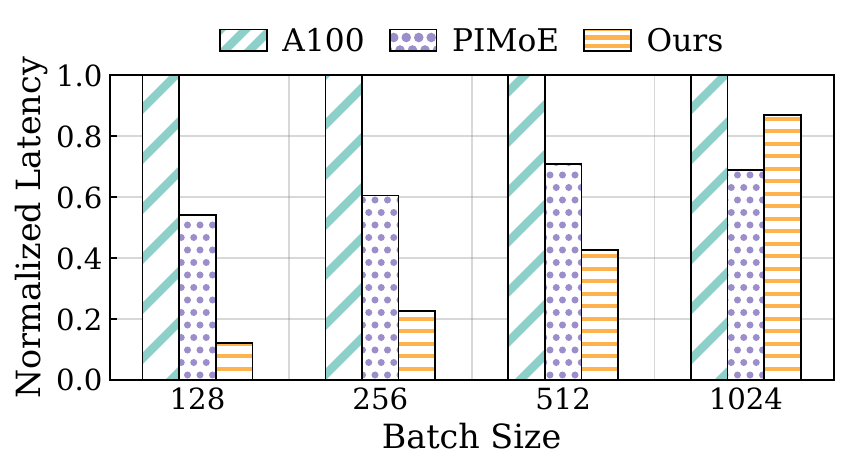}
    \vspace{-12pt}
    \caption{Switch-Base-16.}
    \label{fig:sec5_pimoe_base}
  \end{subfigure}\hfill
  \begin{subfigure}[t]{0.49\columnwidth}
    \centering
    \includegraphics[width=\linewidth]{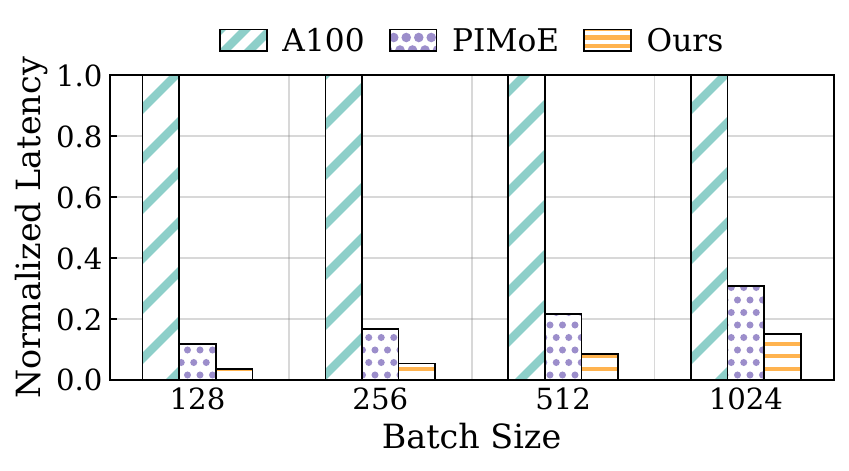}
    \vspace{-12pt}
    \caption{Switch-Large-128.}
    \label{fig:sec5_pimoe_large}
  \end{subfigure}
  \vspace{-6pt}
  \caption{Comparison with PIMoE~\cite{wu2025pimoe} (normalized to A100 GPU baseline).}
  \label{fig:sec5_pimoe}
  \vspace{-12pt}
\end{figure}

\noindent\textbf{MoE FFN Component-level Speedup.} Since PIMoE~\cite{wu2025pimoe} targets MoE FFN acceleration, Fig.~\ref{fig:sec5_pimoe} compares the MoE FFN execution of SPIMOE against both PIMoE and A100 under identical routing decisions. FFN latency includes both computation and communication, excluding latency hidden by pipeline overlap, and is normalized to A100. SPIMOE consistently outperforms both baselines; on Switch-Large-128, it achieves from $3.33\times$ to $2.03\times$ speedup over PIMoE as the batch size increases from 128 to 1024.

\noindent\textbf{Hardware Overhead.}
The synthesized SRAM-PIM subsystem occupies 183.2~mm$^2$ and consumes 0.787~W, while the added compute and buffer logic in HBM-PIM occupies 0.268~mm$^2$ area and consumes 9.599~W. 
Power is evaluated on Qwen3-30B-A3B with batch size 8 and context length 256.
Thermal analysis shows a steady-state HBM-PIM temperature of approximately 46.7$^\circ$C during GEMV execution.

\subsection{Accuracy Evaluation}
\label{subsec:accuracy}



\begin{figure}[t]
  \centering
  \begin{subfigure}[t]{0.57\linewidth}
    \centering
    \includegraphics[width=\linewidth]{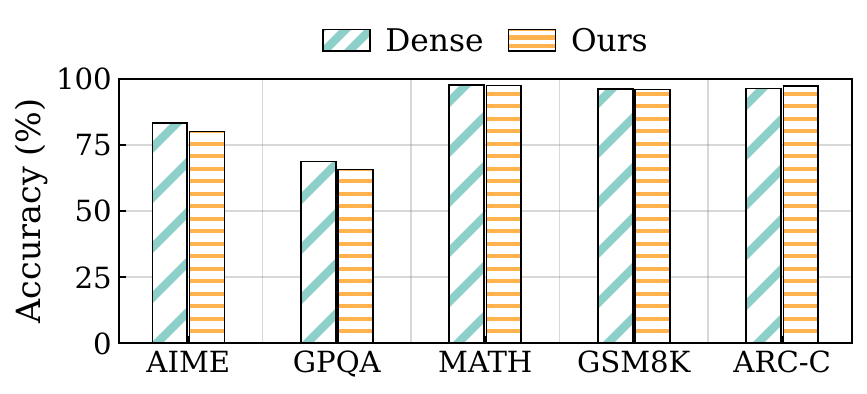}
    \caption{Sparse attention on Qwen3-30B-A3B.}
    \vspace{-6pt}
    \label{fig:sec5_accuracy_sparse_attn_moe_qwen}
  \end{subfigure}\hfill
  \begin{subfigure}[t]{0.4\linewidth}
    \centering
    \includegraphics[width=0.83\linewidth]{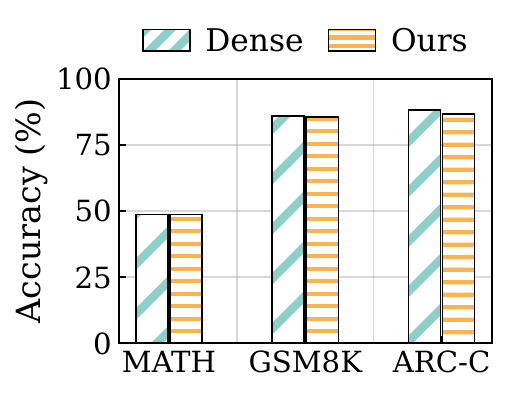}
    \caption{Sparse attention on Phi-mini-MoE-instruct.}
    \vspace{-6pt}
    \label{fig:sec5_accuracy_sparse_attn_moe_phi}
  \end{subfigure}\\[6pt]
  \begin{subfigure}[t]{0.48\linewidth}
    \centering
    \includegraphics[width=\linewidth]{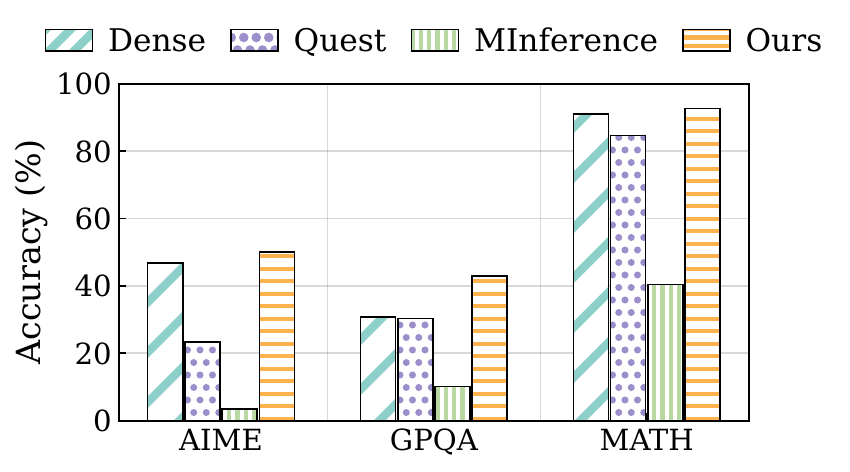}
    \caption{Block sparse attention vs.\ related works on Qwen3-1.7B.}
    \label{fig:sec5_accuracy_sparse_attn_dense}
  \end{subfigure}\hfill
  \begin{subfigure}[t]{0.48\linewidth}
    \centering
    \includegraphics[width=\linewidth]{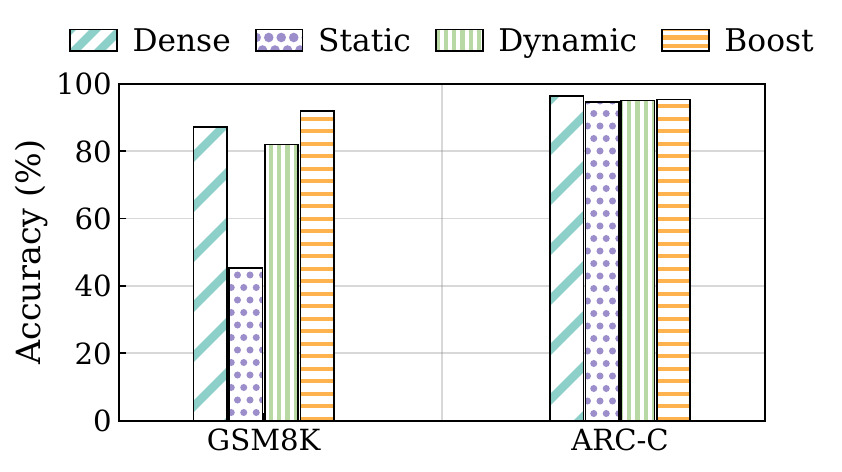}
    \caption{Adaptive expert routing strategies on Qwen3-30B-A3B.}
    \label{fig:sec5_accuracy_expert_routing}
  \end{subfigure}
   \vspace{-6pt}
  \caption{Accuracy evaluation results.}
   \vspace{-12pt}
  \label{fig:sec5_accuracy}
\end{figure}

\noindent\textbf{Block-sparse Attention.}
Fig.~\ref{fig:sec5_accuracy} evaluates the accuracy of block-sparse attention.
Figs.~\ref{fig:sec5_accuracy_sparse_attn_moe_qwen} and~\ref{fig:sec5_accuracy_sparse_attn_moe_phi} show that block-sparse attention preserves reasoning accuracy across two MoE models and five datasets while reducing the KV cache by approximately 50\% through physical eviction. Since Phi-mini-MoE-instruct supports a maximum output length of 4K tokens, AIME and GPQA are omitted for this model.
On Qwen3-1.7B (Fig.~\ref{fig:sec5_accuracy_sparse_attn_dense}), our method maintains accuracy close to full attention and outperforms Quest~\cite{tang2024quest} and MInference~\cite{jiang2024minference}.
We use a sink size of 128 tokens, a local window of 128 tokens, and an 80\% sparsity ratio, and enable block-sparse attention only for sequences longer than 1K tokens to ensure that at least one block is retained.

\noindent\textbf{Adaptive Expert Routing.}
Fig.~\ref{fig:sec5_accuracy_expert_routing} evaluates three expert routing strategies on Qwen3-30B-A3B with a 2K output budget. \emph{Static} applies a global uniform pruning coefficient $\alpha$ across all layers and phases. \emph{Dynamic} employs depth-stratified coefficients with higher minimum expert counts in shallow layers and reduced thresholds during the thinking phase. \emph{Boost} extends the dynamic strategy by additionally boosting the cognitive experts during the thinking phase. The static strategy leads to severe accuracy degradation as it removes experts critical for reasoning without distinction. The cognitive-enhanced strategy preserves reasoning quality and achieves accuracy comparable to the dense baseline, confirming that reinforcing reasoning-critical experts is important under aggressive expert routing.

\subsection{Ablation Studies}
\label{subsec:ablation}


\begin{figure}[t]
  \centering
  \begin{subfigure}[t]{0.51\columnwidth}
    \centering
    \includegraphics[width=\linewidth]{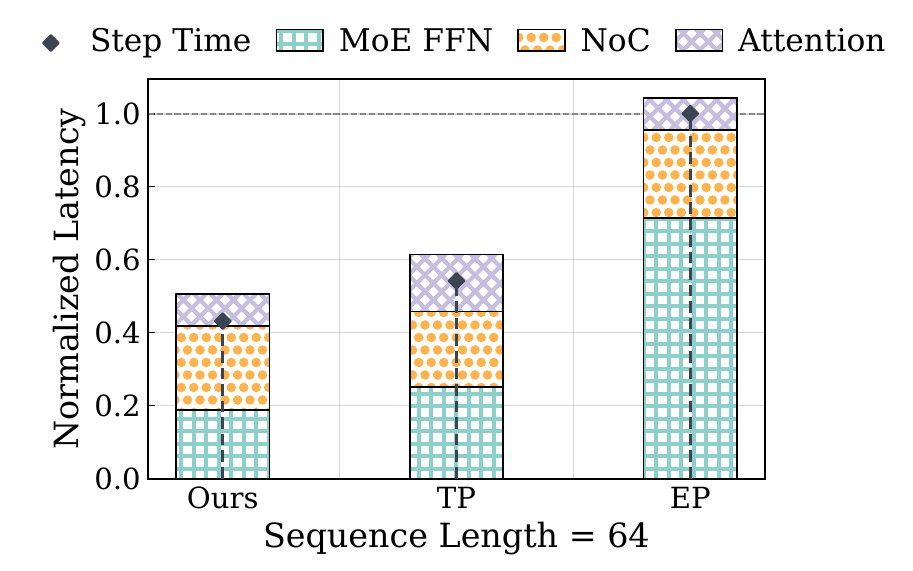}
    \caption{Ablation of expert mapping strategies (normalized to EP baseline).}
    \label{fig:sec5_ablation_placement}
  \end{subfigure}\hfill
  \begin{subfigure}[t]{0.46\columnwidth}
    \centering
    \includegraphics[width=\linewidth]{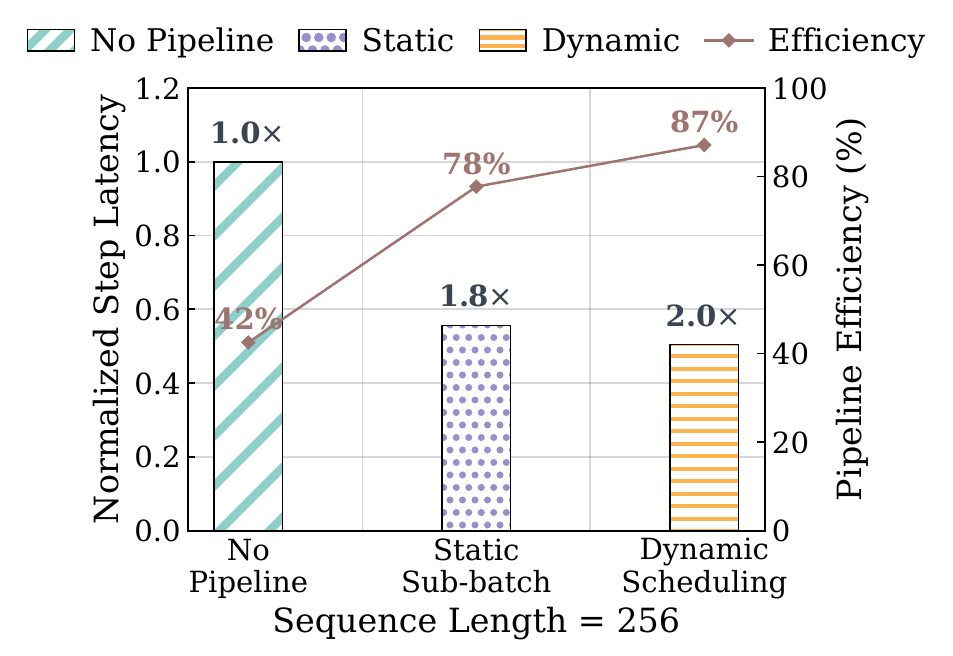}
    \caption{Ablation of dynamic scheduling  (normalized to No Pipeline).}
    \label{fig:sec5_ablation_dynamic}
  \end{subfigure}
  \vspace{-6pt}
  \caption{Ablation studies on expert mapping and dynamic scheduling.}
  \label{fig:sec5_ablation}
  \vspace{-8pt}
\end{figure}

\noindent\textbf{Expert Mapping.}
Fig.~\ref{fig:sec5_ablation_placement} compares expert mapping strategies, with all latencies normalized to the EP baseline. The optimized two-phase TP+EP mapping achieves $1.89\times$ speedup.

\noindent\textbf{Dynamic Scheduling.}
Fig.~\ref{fig:sec5_ablation_dynamic} evaluates the impact of sub-batch pipelining and dynamic scheduling. Compared to a non-pipelined baseline, static sub-batch partitioning achieves $1.80\times$ speedup with $77.7\%$ pipeline efficiency, while the dynamic scheduling further improves to $1.99\times$ speedup and $87.3\%$ pipeline efficiency.

\begin{figure}[t]
\centering
\begin{minipage}{0.9\columnwidth} 
  
  \begin{adjustbox}{valign=m}
    \begin{minipage}{0.39\linewidth} 
      \begin{subfigure}{\linewidth}\captionsetup{justification=raggedright,singlelinecheck=false}
        \centering
        \includegraphics[width=0.6\linewidth]{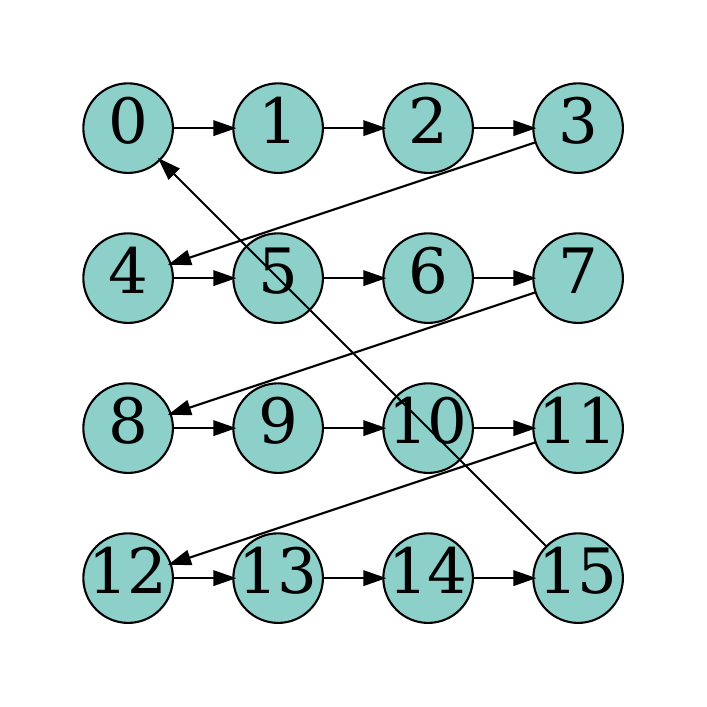}
        \vspace{-6pt}
        \caption{Original topology.}
        \label{fig:sec5_ring_original}
      \end{subfigure}
      \vspace{1mm} 
      \begin{subfigure}{\linewidth}\captionsetup{justification=raggedright,singlelinecheck=false}
        \centering
        \includegraphics[width=0.6\linewidth]{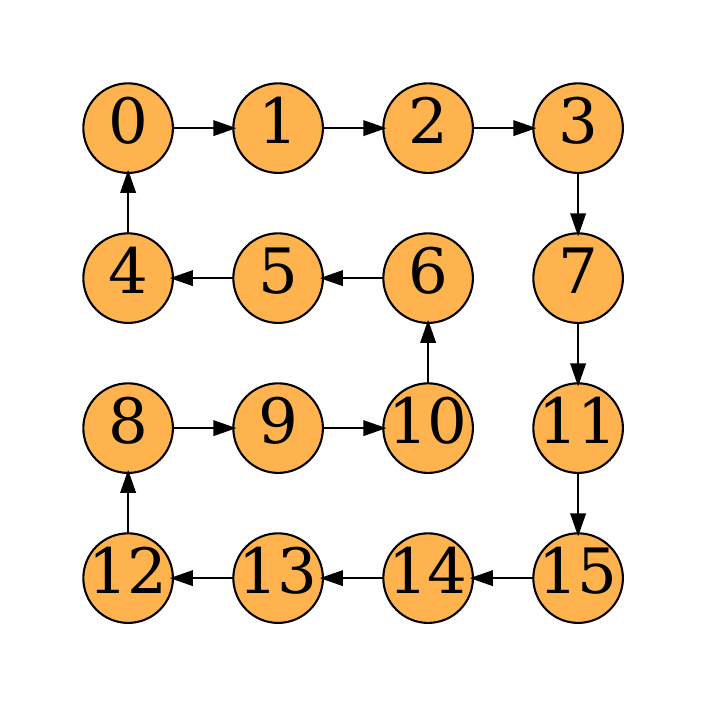}
        \vspace{-6pt}
        \caption{Optimized topology.}
        \label{fig:sec5_ring_optimized}
      \end{subfigure}
    \end{minipage}
  \end{adjustbox}
  \hfill
  \begin{adjustbox}{valign=m}
    \begin{minipage}{0.57\linewidth}
      \vspace{-6pt} 
      \begin{subfigure}{0.8\linewidth}
        \includegraphics[width=\linewidth]{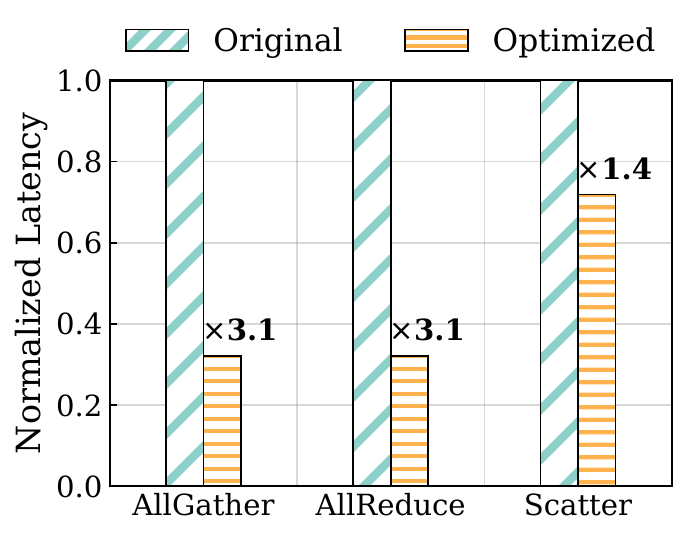}
        \caption{NoC optimization (normalized per operator).}
        \label{fig:sec5_ablation_noc}
      \end{subfigure}
    \end{minipage}
  \end{adjustbox}
  \vspace{-6pt}
  \caption{NoC ring topology comparison and optimization ablation.}
  \vspace{-12pt}
  \label{fig:sec5_ring_compare}
\end{minipage}
\end{figure}

\noindent\textbf{NoC Topology Optimization.}
Fig.~\ref{fig:sec5_ablation_noc} shows the per-operator latency reduction from NoC topology optimization. All-gather and all-reduce operations each achieve $3.12\times$ speedup through conflict-free scheduling on the NoC mesh. The NoC optimization yields a $1.16\times$ overall speedup in the end-to-end test. Fig.~\ref{fig:sec5_ring_compare} visualizes the NoC ring topologies before and after optimization, showing how the NoC topology optimization algorithm redistributes communication paths to eliminate link conflicts and reduce the maximum number of serialized gather rounds.



\begin{figure}[t]
  \centering
  \begin{subfigure}[t]{0.49\columnwidth}
    \centering
    \includegraphics[width=\linewidth]{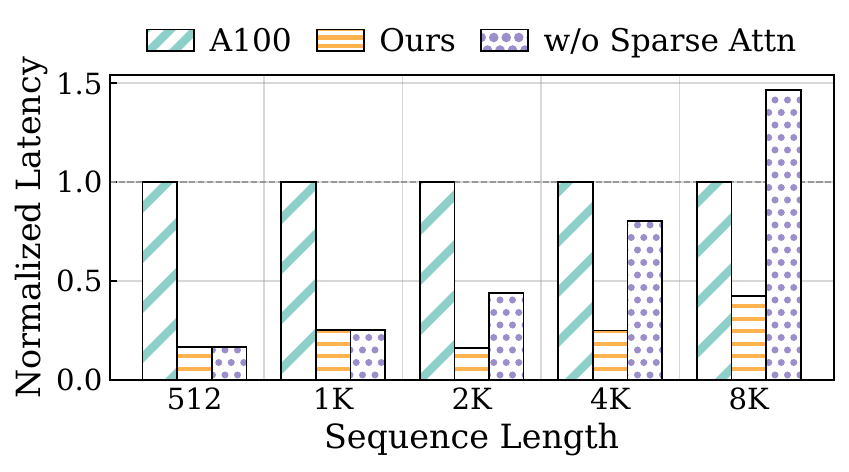}
    \caption{Ablation of sparse attention (Batch Size = 8).}
    \label{fig:sec5_ablation_sparse_attn}
  \end{subfigure}\hfill
  \begin{subfigure}[t]{0.49\columnwidth}
    \centering
    \includegraphics[width=\linewidth]{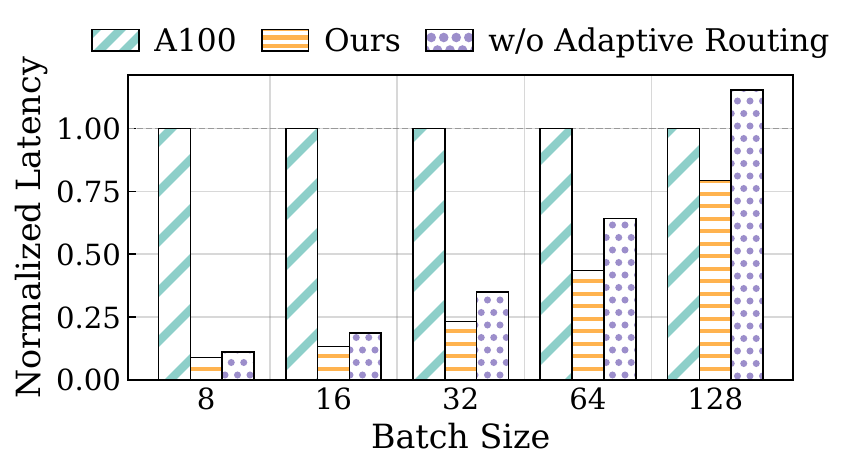}
    \caption{Adaptive routing ablation (Sequence Length = 64).}
    \label{fig:sec5_ablation_routing}
  \end{subfigure}
  \vspace{-6pt}
  \caption{Ablation studies on sparse attention and adaptive expert routing.}
  \vspace{-12pt}
  \label{fig:sec5_ablation_sparsity}
\end{figure}

\noindent\textbf{Sparse Attention.}
Fig.~\ref{fig:sec5_ablation_sparse_attn} evaluates the impact of block sparse attention on end-to-end latency under fixed batch size 8 with varying sequence lengths. As sequence length increases beyond 1K, the sparse attention mechanism progressively reduces the SRAM-PIM workload. At 8K tokens, the system with sparse attention achieves $2.36\times$ speedup, while the system without sparse attention becomes $1.47\times$ slower than the A100 baseline. This demonstrates that block sparse attention is essential for maintaining PIM acceleration at long sequences.

\noindent\textbf{Adaptive Expert Routing.}
Fig.~\ref{fig:sec5_ablation_routing} compares the system with and without adaptive expert routing at short sequence lengths where MoE FFN dominates execution time. Without adaptive routing, all top-K experts are activated, increasing the FFN computation. With adaptive routing, the system selectively prunes redundant experts, achieving a 1.53$\times$ speedup with batch size 128. The performance improvement is observed across different batch sizes, confirming that adaptive expert routing complements the architectural advantage of PIM by reducing the effective computation per token.

%% file: tex/section5-conclusion.tex
\section{Conclusion}
In this paper, we present SPIMOE, the first hybrid-sparse heterogeneous PIM framework for efficient long-chain reasoning inference in MoE models. 
SPIMOE jointly addresses KV cache-intensive attention and highly sparse, irregular expert-FFN execution through reasoning-aware algorithm-hardware co-design. 
At the algorithm level, it integrates adaptive expert routing and block-sparse attention with KV cache eviction to eliminate redundant expert computation, attention computation, and KV-cache storage. Architecturally, SPIMOE realizes Attention-FFN Disaggregation on a heterogeneous PIM architecture integrating SRAM-PIM and HBM-PIM, effectively minimizing both data movement and storage overhead. 
Experimental results show that SPIMOE achieves up to $8.35\times$ end-to-end speedup over an NVIDIA A100 GPU and $3.33\times$ speedup in MoE FFN execution over PIMoE, while maintaining reasoning accuracy comparable to full-attention baselines. These results demonstrate that the synergy between reasoning-aware sparsity and heterogeneous PIM design is a promising direction for scaling future sparse LLM inference systems.